\documentclass[aps,prd,twocolumn,groupedaddress]{revtex4}

\usepackage{graphicx}

\begin{document}

\title{Self-similar spherically symmetric solutions of the massless
Einstein-Vlasov system}

\author{Jos\'e M. Mart\'\i n-Garc\'\i a}
\email[]{J.M.Martin-Garcia@maths.soton.ac.uk}

\author{Carsten Gundlach}
\email[]{C.Gundlach@maths.soton.ac.uk}

\affiliation{
Faculty of Mathematical Studies, University of Southampton,
Southampton SO17 1BJ, UK}

\date{\today}


\begin{abstract}
We construct the general spherically symmetric and self-similar
solution of the Einstein-Vlasov system (collisionless matter coupled
to general relativity) with massless particles, under certain
regularity conditions. Such solutions have a curvature singularity by
construction, and their initial data on a Cauchy surface to the past
of the singularity can be chosen to have compact support in momentum
space.  They can also be truncated at large radius so that they have
compact support in space, while retaining self-similarity in a central
region that includes the singularity.  However, the Vlasov
distribution function can not be bounded. As a simpler illustration of
our techniques and notation we also construct the general spherically
symmetric and static solution, for both massive and massless
particles.
\end{abstract}


\maketitle



\section{Introduction}

In an astrophysical process of gravitational collapse of matter there
is always some physical effect opposing the force of gravity, so that
not all the matter necessarily ends up in the final black hole. Since
the initial work of Choptuik \cite{Choptuik}, it has become clear
that in many physical systems it is possible to fine tune the initial
conditions so that collapse can result in black holes of arbitrarily
small mass. This gives rise to interesting effects which are now
studied as a particular branch (called type II) of Critical Phenomena
in Gravitational Collapse (see \cite{Gundlach_livingreviews} for a
review). In a perfect fluid it is the pressure that opposes
collapse. In field theories, it appears to be the fact that the matter
field propagates at the speed of light.

Here we study another type of matter, a cloud of point particles that
do not interact directly through collisions but only through their
averaged gravitational field. This system is described by the coupled
Einstein and Vlasov equations. In collisionless matter it is the
tendency of particles in the cloud to miss each other that acts as the
source of dispersion. In a spherically symmetric situation, such as we
shall study here, one can also think of their angular momentum as the
force opposing gravity. (In spherical symmetry, each particle has
angular momentum, but the total angular momentum of all particles is
zero.)  

In numerical simulations of the gravitational collapse of
asymptotically flat, spherically symmetric collisionless matter
configurations Rein, Rendall and Schaeffer \cite{ReinRendallSchaeffer}
have not found any sign of critical phenomena at the black hole
threshold. Olabarrieta and Choptuik
\cite{OlabarrietaChoptuik,Olabarrieta} have not found type II critical
phenomena either, but have found some evidence of type I critical
phenomena, where a metastable static solution acts as an intermediate
attractor at the black hole threshold. They have found the expected
scaling of the lifetime of the intermediate attractor with distance to
the black hole threshold. The critical exponent, however, varies from
5.0 to 5.9 between families of initial data, for a quoted uncertainty
of 0.2. The Vlasov distribution function of the critical solution was
not found to be universal, and the metric of the critical solution was
found to be universal only up to a rescaling of space.

We want to explore the problem of existence of critical phenomena in
the Einstein-Vlasov system by looking for both static and continuously
self-similar (from now on CSS) solutions of the system and then
testing for their stability against linear perturbations. A static
solution with exactly one growing mode is a potential critical
solution for type I critical collapse, and a self-similar solution
with exactly one growing mode is a potential critical solution for
type II critical collapse. In this paper, we construct the general
static and the general self-similar solutions, assuming some
regularity conditions. In order to avoid introducing a scale, we
further restrict ourselves to massless particles in the self-similar
case. Their perturbations and the extension to massive particles will
be studied elsewhere. The static solutions of the Einstein-Vlasov
system have been extensively studied already. We deal with them here
partly as an introduction to the methods we use later in the CSS case
but also for the light they shed on type I critical phenomena. To our
knowledge little is known about CSS solutions: The CSS equations were
derived in \cite{Munier}, but no CSS solutions were found. Spherically
symmetric CSS solutions for rotating dust, where $L(r)$ can be
specified freely in the initial data on top of $\rho(r)$ and $v(r)$,
were constructed recently by Gair \cite{Gair}.

We begin by fixing notation and stating the field equations for the
Einstein-Vlasov system in spherical symmetry. We then introduce the idea
that the general solution of the Vlasov equation on a {\it fixed}
spacetime can be given in terms of an arbitrary function of three
variables that are conserved along trajectories and which can be given
as quadratures if the spacetime has an additional continuous symmetry, 
besides spherical symmetry. This is true in particular for both massive
and massless particles in static spacetime, and for massless particles 
in a CSS spacetime.

The further requirement that the resulting stress-energy of the
collisionless matter is {\it consistent} with the assumption of
staticity or CSS reduces the free function of three variables to one
of two variables. We show this first for the static case, and discuss
an interesting subtlety. We also construct some explicit solutions
numerically. This section can be omitted by the reader who is only
interested in the self-similar case, but we feel that it serves as a
useful warmup. 

In discussing the CSS case, we first discuss the conserved quantities,
and classify the possible particle trajectories on a CSS
spacetime. We then find the general solution on a fixed background
spacetime, and reduce it to the general self-gravitating solution. We
again construct some explicit solutions numerically, and discuss the
general features of these solutions.  The implications of our results
for critical phenomena in gravitational collapse are discussed in the
Conclusions section.

\section{The Einstein-Vlasov system in spherical symmetry}

\subsection{Collisionless matter}

In this section we briefly review the essential ideas and equations 
needed in the rest of the Article. For a longer introduction to the 
Einstein-Vlasov system see \cite{Ehlers}. We add some considerations about the 
integration of the Vlasov equation and its relation to the symmetries 
of the spacetime.

The Einstein-Vlasov system describes the evolution of a statistical
ensemble of non-interacting particles coupled to gravity through their
averaged stress-energy. It is possible to work with particles of
different masses, but in this work we restrict ourselves to a single
particle species, with mass $m\ge 0$.  We keep $m$ arbitrary, so that
all the equations in this section are valid both for massless and
massive particles.

As in classical statistical mechanics we describe the state of a many
body system with a positive distribution function over the phase space
of the system. As the particles do not interact except through their
averaged stress-energy, we can use a distribution function $f(x^\mu,
p^\nu)$ on the one-particle phase space, where $x^\mu$ are coordinates
in spacetime and the coordinates $p^\mu$ are constructed using the
metric at the point $x^\mu$ from the coordinates $p_{\mu}$ in its
cotangent space at that point. As we always have $g_{\mu\nu}p^\mu
p^\nu=-m^2$ and $p^0>0$, we can consider the distribution $f$ as a
function $f(x^\mu, p^i)$. Greek indices denote the range 0-3 and Latin
indices the range 1-3.

The free-fall trajectories of the particles define a congruence of
curves on phase space, tangent to the Liouville operator (or
``geodesic spray'')
\begin{equation}
{\cal L}
  = \frac{d}{d\sigma}
  = \frac{dx^\mu}{d\sigma} \frac{\partial}{\partial x^\mu}
  + \frac{dp^i}{d\sigma} \frac{\partial}{\partial p^i}
  = p^\mu \frac{\partial}{\partial x^\mu} -
  \Gamma^i_{\nu\lambda} p^\nu p^\lambda \frac{\partial}{\partial p^i}.
\end{equation}
In this article $\sigma$ is an affine parameter related to proper time
$s$ through $ds=m\, d\sigma$ for massive particles. This allows us to
discuss both massive and massless particles simultaneously. As the
particles do not interact directly, the evolution of $f$ is governed
by the Vlasov equation
\begin{equation}
{\cal L}f=0 .
\label{Vlasov}
\end{equation}

The stress-energy tensor at each point is obtained by integrating the
distribution function over momentum space:
\begin{equation}
\label{Tmunu}
T^{\mu\nu}(x) = 
\int_{P(x)} \frac{d^3p}{-p_0} \sqrt{-g} f(x,p) p^\mu p^\nu,
\end{equation}
where $P(x)$ is the 3-momentum space at the point $x^\mu$ and $p_0$ is
determined from $p^i$ and the metric. The Vlasov equation is a
sufficient condition for stress-energy conservation. The particle
number current given by
\begin{equation}
\label{Nmu}
N^{\mu}(x) = \int_{P(x)} \frac{d^3p}{-p_0} \sqrt{-g} f(x,p) p^\mu .
\end{equation}
is also conserved. Note that the distribution function $f(x,p)$ is a
scalar on phase space even though there is a factor $\sqrt{-g}$ in the
integrals.  The natural volume measure on the phase space is
$-dx^{0123}\wedge dp_{0123}$, which can be rewritten as $(\sqrt{-g}\,
dx^{0123})\wedge(\sqrt{-g}\, dp^{0123})$. It is the second factor
$\sqrt{-g}$ that appears in the integrals (\ref{Tmunu}) and
(\ref{Nmu}), which are themselves tensors on spacetime and would
therefore be integrated over spacetime using the measure $(\sqrt{-g}\,
dx^{0123})$.

A comment on the dimensions of the variables. It would seem natural to
measure the mass $m$ and energy-momentum $p$ of individual particles
in the same units one measure gravitational mass and energy-momentum
in, for example in the stress-energy tensor. But particle momentum
appears in the stress-energy momentum tensor only under the
integration over momentum space. Dimensional analysis is therefore
less restrictive than one might assume. 

It is consistent to assign particle mass and energy-momentum a
dimension $P$ that is independent of the dimension $M$ of
gravitational mass. In units in which $c=G=1$, the unit of
gravitational mass $M$ is equal to the unit of length $L$, but $P$
remains independent. For example, $m$ and $p^\mu$ have dimension $P$,
and angular momentum squared $F=|\vec{x}\wedge\vec{p}|^2$ has
dimensions $L^2P^2$, but the stress-energy tensor has dimension
$L^{-2}$, which can be thought of as $ML^{-3}$.  The distribution
function relates both kinds of masses and has dimensions
$L^{-2}P^{-4}$. The particle current has dimension $P^{-1}L^{-2}$,
which can be thought of as $MP^{-1} L^{-3}$: it has the dimension
$L^{-3}$ of a number density only if gravitational and particle mass
are measured in the same units.

\subsection{Spherical symmetry}

Now we impose spherical symmetry. We describe the
metric using polar-radial coordinates,
\begin{equation}
ds^2=-\alpha^2(t,r) dt^2+a^2(t,r)dr^2
     +r^2(d\theta^2+\sin^2\theta d\phi^2).
\end{equation}
In this coordinate choice there is still the gauge freedom $t\to
t'(t)$, which changes the lapse $\alpha$. We fix this freedom by
setting $\alpha=1$ either at $r=0$ or at $r=\infty$.

The Einstein equations give the following equations for the first
derivatives of the metric coefficients:
\begin{eqnarray}
\frac{\alpha_{,r}}{\alpha}&=&
\frac{a^2-1}{2r}+\frac{ra^2}{2}8\pi{T_r}^r, \label{alphar} \\
\frac{a_{,r}}{a}&=&-\frac{a^2-1}{2r}-\frac{ra^2}{2}8\pi{T_t}^t, 
\label{ar} \\
\frac{a_{,t}}{a}&=&\frac{ra^2}{2}8\pi{T_t}^r.
\end{eqnarray}
The fourth Einstein equation, involving $T_{\theta\theta}$, is a
combination of derivatives of these three. The third of these
equations, for $a_{,t}$, becomes an identity when the other two
equations are obeyed at each $t$. 

In spherical symmetry, the distribution function has to be of the form
$f(t,r,p^r,|p|)$, where $|p|^2\equiv g_{ij}p^i p^j$. We can simplify the 
Vlasov equation using the fact that angular momentum is a constant of 
motion. Following \cite{Rein95} we define the variables
\begin{eqnarray}
w &\equiv& a p^r, \\
F &\equiv& r^2 (|p|^2-w^2) = r^4 ({p^\theta}^2+\sin^2\theta {p^\phi}^2).
\end{eqnarray}
We also introduce the shorthand
\begin{equation}
W\equiv \sqrt{m^2+w^2+F/r^2} =\alpha p^t.
\end{equation}
In these variables the Vlasov equation in a spherically symmetric
spacetime becomes
\begin{equation}
\frac{\partial f}{\partial t}
+
\frac{\alpha}{a} \frac{w}{W}
\frac{\partial f}{\partial r}
+
\left(
\frac{F \alpha}{r^3 a W}
- W \frac{\alpha_{,r}}{a}
- w \frac{a_{,t}}{a}
\right)
\frac{\partial f}{\partial w}
=0 ,
\end{equation}
where $f$ is now a function $f(t,r,w,F)$. There is no derivative
with respect to $F$ because it is a constant of motion.

The nonvanishing components of the particle number current are
\begin{eqnarray}
N^t &=&\frac{\pi}{r^2 \alpha} \int_0^\infty dF \int_{-\infty}^\infty 
dw \; f , 
\label{Nt} \\
N^r &=& \frac{\pi}{r^2 a} \int_0^\infty dF \int_{-\infty}^\infty 
dw \; f \frac{w}{W} ,
\label{Nr}
\end{eqnarray}
The distribution function is always nonnegative and therefore we have
$\alpha N^t\ge a|N^r|$, which means that $N^\mu$ is always timelike or
lightlike. It is lightlike only for distributions where all particles
are massless and moving radially ($F=0$) in the same direction, so
that either $w=W$ or $w=-W$.
The non-vanishing components of the stress-energy tensor are
\begin{eqnarray}
T_r{}^r &=& \frac{\pi}{r^2} 
\int_0^\infty dF \int_{-\infty}^\infty dw
\; f \; \frac{w^2}{W} \ge 0, 
\label{Trr} \\
T_t{}^t &=& -\frac{\pi}{r^2} 
\int_0^\infty dF \int_{-\infty}^\infty dw
\; f \; W \le 0, 
\label{Ttt} \\
T_t{}^r &=& -\frac{\pi}{r^2}\frac{\alpha}{a} 
\int_0^\infty dF \int_{-\infty}^\infty dw
\; f \; w ,
\label{Ttr} \\
{T_\phi}^\phi  = 
{T_\theta}^\theta &=& \frac{\pi}{2r^4 } 
\int_0^\infty dF \int_{-\infty}^\infty dw
\; f \; \frac{F}{W} \ge 0.
\label{Tphiphi}
\end{eqnarray}
This tensor is conserved and satisfies the dominant and strong energy
conditions. We assume that $f$ behaves in such a way that all
integrals (\ref{Nt}-\ref{Tphiphi}) converge and are finite at every
point. 
 
Assuming $a\ge 1$ the following inequalities follow from the Einstein
equations:
\begin{eqnarray}
\alpha_{,r} &\ge& 0, \\
(a\alpha)_{,r} &\ge& \left|(a^2)_{,t}\right| , \\
\frac{\alpha}{a}\left(\frac{a}{\alpha}\right)_{,r} &\ge& 
\frac{1-a^2}{r} .
\end{eqnarray}
Finally we have that
\begin{equation}
T_\mu{}^\mu = -m^2 \frac{\pi}{r^2} 
\int_0^\infty dF \int_{-\infty}^\infty dw 
\; \frac{f}{W} \le 0 ,
\end{equation}
and therefore the Ricci scalar $R$ is always nonnegative. For plotting
results we shall also use the Hawking (or Misner-Sharp) mass function
$M(t,r)$, which is defined by
\begin{equation}
a^{-2}(t,r)\equiv 1-{2M(t,r)\over r}.
\end{equation}

\subsection{Further symmetries and conserved quantities}

The Vlasov equation (\ref{Vlasov}) expresses the fact that $f$ is
constant along trajectories of particles. Therefore the solution of
that equation is, formally, an arbitrary function of the constants of
motion of the problem. In the general case $f$ would be a function of
eight constants of motion, but when we impose spherical symmetry $f$
can only depend on some of these. Spherical symmetry divides the four
degrees of freedom of a point particle into two radial and two angular
degrees of freedom. The angular part gives rise to four constants of
motion: the three components $L_x$, $L_y$, $L_z$ of the angular
momentum vector, and the initial angle with respect to the axis given
by this vector. If the averaged stress-energy of the particles
is to be spherically symmetric, however, the distribution function $f$
can only depend on the modulus (squared) $F$ of the angular
momentum. As we saw in the previous subsection, this can be used to 
reduce the Vlasov and Einstein equations to spherical symmetry.
In the same way, additional symmetries can be used to further simplify 
the equations.

The reduced, radial, system has two degrees of freedom $[t(\sigma),
r(\sigma)]$ with conjugate momenta $[p_t(\sigma), p_r(\sigma)]$, where
$\sigma$ is an affine parameter along the particle trajectories. The
Hamiltonian of the reduced system is
\begin{equation}
H=\frac{1}{2}\left(
-\frac{p_t^2}{\alpha^2}
+\frac{p_r^2}{a^2}
+\frac{F}{r^2}
\right) ,
\end{equation}
where $F$ is now a given constant. This Hamiltonian system has four
independent constants of motion $c_1,...,c_4$. One of these is the value
of the Hamiltonian itself, $H= -\frac{m^2}{2}$. The general solution of 
the equations of motion can formally be written as
\begin{equation}
\left.
\begin{array}{rcl}
t(\sigma) &=& t(\sigma,c_1,...,c_4,F) \\ r(\sigma) &=&
r(\sigma,c_1,...,c_4,F) \\ p_t(\sigma) &=& p_t(\sigma,c_1,...,c_4,F)
\\ p_r(\sigma) &=& p_r(\sigma,c_1,...,c_4,F)
\end{array}
\right\} \quad\Rightarrow\quad
f=f(c_1,...,c_4,F) ,
\end{equation}
where we now consider the constants $c_i$ as functions of $t$, $r$,
$p_t$, $p_r$ and $\sigma$. Let $c_4=m$. In the following we do not
refer to $m$ as a conserved quantity, but consider it a parameter of
the equations. At least one of the four constants of motion, say
$c_3$, must depend on $\sigma$, but because $f$ does not depend
explicitly on $\sigma$, $f$ can not depend on this constant of
motion. Therefore, the general solution $f$ in spherical symmetry will
be an arbitrary function of $F$ and two nontrivial constants of motion
$c_1$ and $c_2$.

Although the two nontrivial constants of motion $c_1$ and $c_2$ always
exist formally (and we can find them by integrating the equations of
motion numerically), finding their analytical expressions in terms of
the variables of the problem is only possible when the reduced radial
system is integrable. Using the Liouville theorem \cite{Arno89}, 
we need to find
only one constant of motion in terms of $(t,r,p_t,p_r)$, because then
we will have two constants in involution for a system with two degrees
of freedom. The easiest way of finding such a constant is imposing an
additional continuous symmetry.  In the following we analyze two
possible symmetries: staticity and self-similarity.  We will then be
able to give analytic expressions (at least quadratures) for the
trajectories of the particles, simplifying the calculation of the
energy-momentum tensor components. This is equivalent to giving a
formal solution of the Vlasov equation in terms of its characteristics
in phase space.

Even though we are interested in self-similarity, we first review
the static case to show some important ideas in a well-known context.

\section{Review of static spacetimes} 

\subsection{Field equations} 

A static spacetime has an additional Killing vector
\begin{equation}
\xi=\partial_t ,
\end{equation}
so that the metric functions $\alpha$ and $a$ are just functions of
$r$.  The particles have an additional constant of motion
\begin{equation}
E\equiv-\xi^\mu p_\mu=- p_t=\alpha W=\alpha\sqrt{m^2+w^2+F/r^2}.
\end{equation}
The four constants of motion (of the full 4-dimensional system) $p^2$,
$F$, $L_z$ and $E$ are in involution and therefore, by the Liouville
theorem, we can integrate the system. The radial equations of motion
in terms of the constants of motion $E$ and $F$ are
\begin{eqnarray}
\frac{dr}{d\sigma} &=& 
p^r = \pm
\frac{1}{a(r)} 
\sqrt{ \frac{E^2}{\alpha(r)^2}-m^2-\frac{F}{r^2}}, \\
\frac{dt}{d\sigma} &=& 
p^t =
\frac{E}{\alpha^2(r)} .
\end{eqnarray}
Defining $v \equiv p^r/p^t$ we can eliminate $\sigma$ and integrate the 
equations to obtain the quadrature
\begin{equation}
\label{t0}
t-t_0 = \int_{r_0}^{r(t)} \frac{dr'}{v(r')} ,
\end{equation}
where
\begin{equation}
{dr\over dt}\equiv v(r) = \pm \frac{\alpha(r)}{a(r)}
\sqrt{ 1 - \frac{m^2\alpha^2(r)}{E^2} - \frac{\alpha^2(r)F}{r^2E^2} } .
\end{equation}
We have the new constant of motion
\begin{equation}
\label{t0constant}
t_0(t,r,E,F)=t-\int^r_{r_0} \frac{dr'}{v(r')},
\end{equation}
which is the time when the particle will be at $r_0$, given that it is
at $r$ at time $t$. After $E$, this is the second of the nontrivial
constants of motion $c_1$ and $c_2$ referred to in the general
discussion.

Therefore, the most general solution to the Vlasov equation on a
static background can be given as an arbitrary function of three
variables
\begin{equation}
\label{Vlasovstatic}
\hbox{Vlasov in static spacetime} \quad \Rightarrow \quad
f(t,r,w,F)=g\left(t_0, E, F\right) .
\end{equation}
As $t_0$ depends explicitly on $t$, this allows for a solution $f$
that is explicitly time-dependent. 

If particles with a given $E$ and $F$ have bound orbits, $g$ must be
periodic in $t_0$ with the period of that orbit. The period is
determined as the integral (\ref{t0}) between the two turning points,
and depends on $E$ and $F$. We can avoid this complicated periodicity
of $g(t_0,E,F)$ by inverting equation (\ref{t0constant}) to obtain a
constant of motion $r_0(t,r,E,F)$ which gives the position $r_0$ of a
particle with $E,F$ at some canonical time $t_0$ given that it is at
$r$ at time $t$:
\begin{equation}
\hbox{Vlasov in static spacetime} \quad \Rightarrow \quad
f(t,r,w,F)=\tilde g\left(r_0, E, F\right) .
\end{equation}
This means that we can freely specify $f(r,w,F)$ at $t=t_0$, and still
obtain an automatic solution of the Vlasov equation.

So far we have considered Vlasov matter moving on a given spacetime.
But to be consistent with staticity, we need the stress-energy
momentum to be independent of $t$. Therefore, in principle, $f$ should
not depend on $t_0$ or $r_0$, which depend on $t$. The most general
self-consistent static solution of Einstein-Vlasov in spherical
symmetry would then have the form
\begin{equation}
\hbox{Einstein-Vlasov$+$ staticity} 
\quad \text{`` }\Rightarrow\text{ ''} \quad
f(t,r,w,F)=h(E,F) .
\end{equation}
This result is referred to as Jeans' theorem and it is known to
hold in Newtonian physics (that is, in the Vlasov-Poisson 
system), but there are some counterexamples in the Einstein-Vlasov
system. That is why we have put the implication arrow in quotes. In 
order to understand this subtlety we first give some explicit solutions
where that result is valid.

\subsection{Some explicit solutions} 

We choose a function $h$ of two variables. The spacetime can
then be determined by solving the Einstein equations, which become a
system of integral-differential equations. The integrals occur in
forming the stress-energy tensor, which depends on the metric as well as
on the arbitrary function $h(E,F)$.

In a static spherically symmetric solution we have two nontrivial
Einstein equations. They are just the Einstein equations
(\ref{alphar}) and (\ref{ar}) in the general case:
\begin{eqnarray}
{\alpha'\over \alpha}&=&{a^2-1\over2r}+{4\pi^2 a^2\over r} \times 
\nonumber \\
&& \qquad \int_0^\infty dF \int_{-\infty}^\infty dw \; h(\alpha W,F) \; {w^2\over
W}, \label{bgstatic1} \\
{a'\over a}&=&-{a^2-1\over2r}+{4\pi^2 a^2\over r} \times 
\nonumber \\
&& \qquad \int_0^\infty dF \int_{-\infty}^\infty dw \; h(\alpha W,F) \; W, 
\label{bgstatic2}
\end{eqnarray}
where $\alpha$ and $a$ are now functions of $r$ only. In order to carry
out the integrations, it is useful to change the integration variable
from $w$ and $F$ to $E$ and $F$, the arguments of the free function
$h$:
\begin{eqnarray}
{\alpha'\over \alpha}&=&{a^2-1\over2r}+{4\pi^2 a^2\over r\alpha^2} 
\times \nonumber \\
&& \int_0^\infty dF \int_{E_{\rm min}(r,F)}^\infty 2E\,dE \; h(E,F) \;
K^{1/2}
\label{integralalpha} \\
{a'\over a}&=&-{a^2-1\over2r}+{4\pi^2 a^2\over r\alpha^2} \times 
\nonumber \\
&& \int_0^\infty dF \int_{E_{\rm min}(r,F)}^\infty 2E\,dE \; h(E,F) \;
K^{-1/2},
\label{integrala}
\end{eqnarray}
where we have used the shorthand expressions
\begin{eqnarray}
E_{\rm min}(r,F)&\equiv&\alpha(r)\left(m^2+{F\over r^2}\right)^{1/2},
\label{kerneldef} \\
K(r,\alpha,E,F)&\equiv&{w^2\over W^2}=
1-{m^2\alpha^2(r)\over E^2}-{F\alpha^2(r)\over r^2E^2} \nonumber \\
&=& 1-{E_{\rm min}^2\over E^2}
\end{eqnarray}
for the boundary and integral kernel. $E_{\rm min}(r,F)$ gives the
minimum energy of a particle at $r$ with angular momentum $F$, which
corresponds to the state with $w=0$. Note that it is always positive.
It also functions as an effective potential for the radial particle 
motion, as we have
\begin{equation}
v^2={\alpha^2\over a^2E^2}(E^2-E_{\rm min}^2).
\end{equation}

When the function $h(E,F)$ has been specified, the stress-energy
integrals become given functions of $r$ and $\alpha(r)$. The Einstein
equations then become two coupled ordinary differential equations (ODEs)
that determine $a(r)$ and
$\alpha(r)$. Conversely we can think of $a(r)$ and $\alpha(r)$ as
given functions. We then have an integral equation for $h(E,F)$. By
function counting, this integral equation does not determine $h(E,F)$
uniquely. There is an infinite number of functions $h(E,F)$ that give
rise to the same static spacetime, with the same stress-energy tensor.
The change of integration variables to $E$ and $F$ shows that, for
given $h(E,F)$, the integrals in (\ref{integralalpha}) and
(\ref{integrala}) are functions only of the two combinations
$r/\alpha$ and $m\alpha$. In particular, when $m=0$, those integrals
for fixed $h(E,F)$ are functions only of the combination $r/\alpha$.

The particular class of functions $h(E,F)$ where
$h(E,F)=\theta(E_0-E)\theta(F-F_0)\phi(E)(F-F_0)^l$ (called a
``polytrope''), with upper cutoff $E_0$ and lower cutoff $F_0$, and
$l>-1/2$, is analyzed in detail in \cite{Rein94} for massive
particles, where it is shown that solutions exist that describe matter
distributions with finite total mass and compact radial support.

If $h(E,F)$ is restricted to a function of $E$ alone, one finds by
direct comparison of the integrals in the stress-energy tensor that
${T_r}^r={T_\theta}^\theta={T_\varphi}^\varphi$, so that the pressure
is isotropic. This means that the stress-energy content in this case
(spherically symmetric and static) has perfect fluid form. If
$\rho=-{T_t}^t$ and $p={T_r}^r$ are monotonic, one could also read off
a, fairly meaningless, formal ``equation of state''.

It is interesting to solve a simple example in closed form, namely the
ansatz $h(E,F)=c \,\theta(E-E_0)$, with $c$ and $E_0$ positive
constants. This is a ``polytrope'' with $l=0$ and no cutoff in angular
momentum. For convenience of notation we define the dimensionless
constant $\bar m\equiv m/E_0$, and the constant $\bar c \equiv
cE_0^4$. With the gauge choice $\alpha(r=0)=1$ we have $E\ge m$, and
therefore $\bar m$ must be in the range from 0 to 1. The parameter 
$\bar c$ can always be set to 1 by a choice of length
units, and is therefore trivial.  The integrals in equations
(\ref{integralalpha}) and (\ref{integrala}) can be evaluated
explicitly, and give
\begin{widetext}
\begin{eqnarray}
\rho &=& -T_t{}^t =
\frac{\pi \bar c}{2\alpha^4}
\left[
   \sqrt{1-\bar m^2\alpha^2}(2-\bar m^2\alpha^2)
 + \bar m^4\alpha^4
       \log\left(\frac{\bar m\alpha}{1+\sqrt{1-\bar m^2\alpha^2}}\right)
\right],
\label{rhoexact}
\\
p &=& T_r{}^r =
\frac{\pi \bar c}{2\alpha^4}
\left[
   \frac{1}{3}\sqrt{1-\bar m^2\alpha^2}(2-5\bar m^2\alpha^2)
 - \bar m^4\alpha^4
       \log\left(\frac{\bar m\alpha}{1+\sqrt{1-\bar m^2\alpha^2}}\right)
\right] ,
\label{pexact}
\end{eqnarray}
\end{widetext}
for $\bar m\alpha\le 1$, and $\rho=p=0$ otherwise. For massless
particles, these expressions simplify to
\begin{equation}
\rho=3p=\frac{\pi \bar c}{\alpha^4},
\end{equation}
so that we have the formal equation of state $p=\rho/3$. $\alpha(r)$
is an increasing function, defining the surface of the ``star'' at $\bar
m\alpha=1$. Note that lighter particles give larger stars.

\begin{figure*}[ht!]
\begin{center}
\includegraphics[width=8cm]{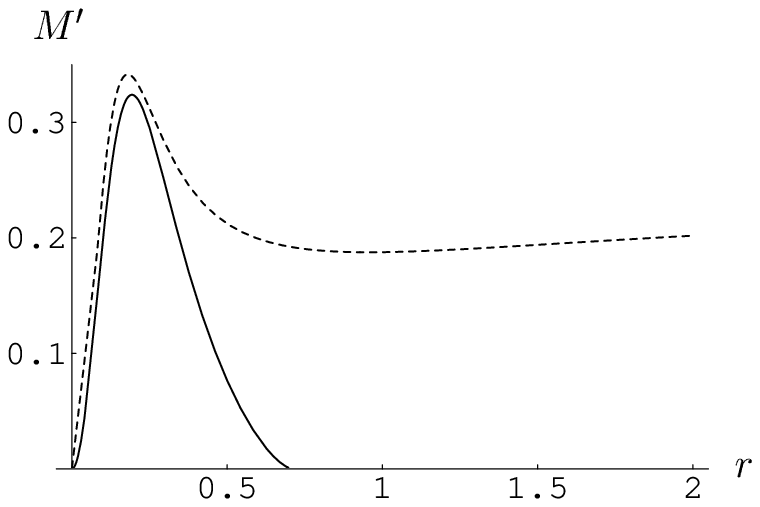}
\includegraphics[width=8cm]{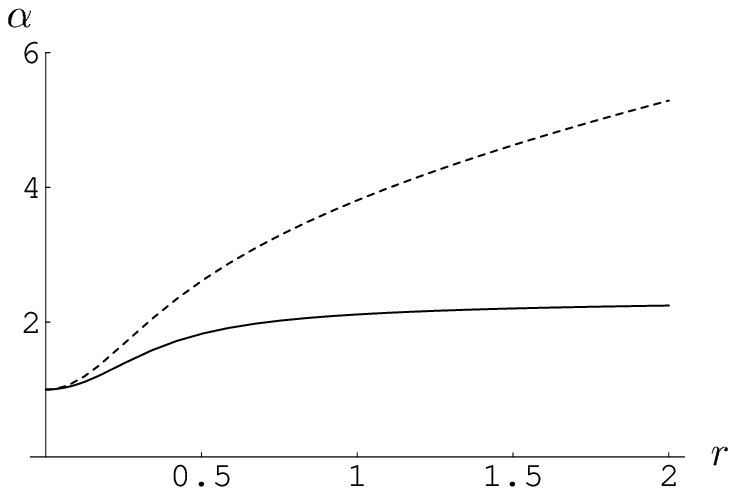}
\caption{ \label{fig:static} Mass distribution $M'(r)$ (left) and lapse
function $\alpha(r)$ (right) corresponding to the solution given in
Eqs. (\ref{rhoexact}) and (\ref{pexact}) for $\bar m=1/2$ (continuous
line) and $\bar m=0$ (dashed line).}
\end{center}
\end{figure*}

An explicit numerical example for the resulting metric is shown in
Fig. \ref{fig:static}, both for massive ($\bar m=1/2$, continuous
line) and massless ($\bar m=0$, dashed line) particles. As we
expected, the massive case gives a star of finite mass and finite
radius, while the massless case gives an infinite mass, infinite size
distribution of matter.  This can be easily understood in terms of the
motion of the particles: in this example massive particles follow
bounded orbits, so that we have a static situation with a finite
number of particles in the system, but massless particles follow
unbounded orbits, so that we need an infinite number of them to get a
static distribution. 

\subsection{Counterexamples to Jeans' Theorem} 
\label{counterexamplesJeans}

We shall now review some static solutions that do, after all, depend
on the third constant of motion $r_0$. Assume a static distribution of
matter and its corresponding static gravitational field. Two different
test particles can have the same energy $E$ and angular momentum $F$
at different values of $r$. It has been proved that in Newtonian
physics (the Vlasov-Poisson system) the distribution function must be
the same at both points for those values of $E$ and $F$ \cite{BFH86}.
This result is referred to as Jeans' theorem. We can understand it
physically in the following way. In Newtonian physics it is always
possible to find a trajectory for one of the particles on which it
reaches the position of the other particle without changing $E$ or
$F$. Because $f$ is constant along particle trajectories it must be
the same at both positions. This result is not true in general
relativity: it is possible to show that two particles with the same
$E$ and $F$ can be separated by a potential barrier. $f$ can then be
consistently different at those positions, and therefore it is not
just a function of $E$ and $F$, but of $r_0$ as well.

The key observation is that the Newtonian minimum energy
\begin{equation}
E_{\rm min}(r,F) = U(r)+\frac{F}{2mr^2},
\end{equation}
where the gravitational potential $U(r)$ obeys the Poisson equation,
always has a single minimum and no maxima, while the relativistic
function may have several extrema.

\begin{figure*}[ht!]
\begin{center}
\includegraphics[width=8cm]{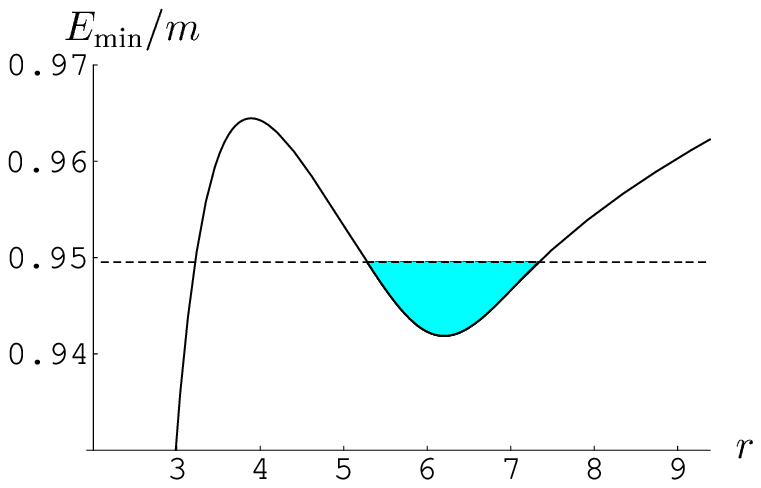} 
\includegraphics[width=8cm]{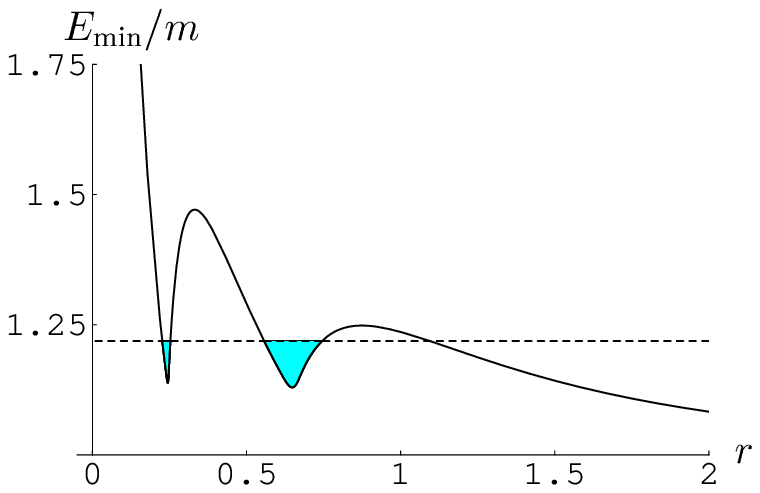} 
\caption{\label{ReinSchaefferfigs} Graphs of the effective potential
$E_{\rm min}(r,F_0)=\alpha(r)\sqrt{m^2+F_0/r^2}$, where $F_0$ is a
lower cutoff in angular momentum. The horizontal dashed line gives the
upper cutoff $E_0$ in energy. We can have particles between both lines
(shaded regions). In both examples, $m=1$, and $h$ is independent of
$E$ and $F$ for $F>F_0$ and $E<E_0$, but takes different values in
different potential wells. In both examples we have set $\alpha=1$ at
$r=\infty$, and so $E_{\rm min}/m\to1$ from below as $r\to\infty$. In
the first graph we have a black hole with $M=1$ and we plot just the
exterior region. $F_0=17$. We could have particles in the region
$2M<r<3M$, but not a static distribution of them. In the second graph
we have two shells of matter, and $F_0=2$. Both of them contain
particles, but $h$ is half in the outer well of what it is in the
inner well. }
\end{center}
\end{figure*}

Rein has given a nice example in \cite{Rein94}: Given a Schwarzschild
black hole of mass $M$, it is impossible to set up a static,
finite-mass distribution of Vlasov particles very close to it (for
$r<3M\dots 6M$, depending on their angular momentum) because it
is not possible to have bound stable orbits in that region. Therefore
we must set $h=0$ in that region. However we can have a static
distribution of particles with the same energy and angular momentum
further away from the black hole. The two regions are separated by a
potential barrier. Therefore $h$ may be different from zero for the
same values of $E$ and $F$. Again, $h$ depends on $r_0$, as well as on
$E$ and $F$.  The left side of Fig. \ref{ReinSchaefferfigs} gives an
explicit numerical example of a self-gravitating shell of
collisionless matter surrounding a black hole.

A second example has been given by Schaeffer \cite{Scha99}.  He shows
that a shell of matter generates a well in the potential which can
sustain another shell, if the inner shell is concentrated near its own
Schwarzschild radius. We have an example in Fig.
\ref{ReinSchaefferfigs} on the right. The second shell can have a
different value of the distribution function $h$ for particles with
the same energy and angular momentum. In the example in Fig.
\ref{ReinSchaefferfigs} $h$ in the second shell is half of $h$ in the
first shell. The same values of $E$ and $F$ are possible in a third
region stretching to infinity (unbound particles), and $h$ has been
chosen to vanish there.

We now state the Einstein-Vlasov form of ``Jeans' theorem''.  In
general relativity particles with given values of
$E$ and $F$ can exist in disjoint regions $n=1,2,...$ for the
particles.  Each region is bounded by turning points
$r^{(n)}_{\pm}(E,F)$. We assume that $r^{(n)}_+<r^{(n+1)}_-$.  If the
first allowed region contains the center then $r^{(1)}_-\equiv 0$ and
if the last allowed region is not bounded then $r^{(n_{\rm
last})}_+\equiv\infty$. The general form of the distribution is
\begin{equation}
\label{GRJeans}
h(r_0,E,F)=
\sum_n \theta(r_0-r^{(n)}_-)\;\theta(r^{(n)}_+-r_0)\;h^{(n)}(E,F).
\label{genJeans}
\end{equation}
In order to find self-gravitating solutions of this form, we
shall usually need a convergent iterative procedure. For example the
shells in the previous counterexample given by Schaeffer can be
constructed iteratively using the fact that in spherical symmetry the
interior shells are not affected by the exterior shells (Birkhoff
theorem).

The dependence of $h$ on $r_0$ in (\ref{GRJeans}) is admittedly rather
restricted, but we have discussed it in some detail, because the
dependence of the equivalent of $h$ in self-similar solutions on all
three constants of motion will be non-trivial, and in fact no
solutions can be found that depend only on $F$ and the equivalent of
$E$: solutions must also depend on the equivalent of $r_0$.

\subsection{Massless particles} 

Massless particles are not very interesting when one looks for static
solutions because it is difficult to form asymptotically flat solutions,
but the system with massless particles has an additional symmetry: the
trajectory of a massless particle, in a given spacetime, for a given
initial position, is completely determined by the direction of its
initial four-momentum, independently of the modulus. In order to make
a given contribution to the stress-energy tensor, we can therefore use
$N$ particles of four-momentum $p^\mu$ or one particle of
four-momentum $Np^\mu$.  We shall see that this allows us to state
explicitly which part of the Vlasov function $h(E,F)$ is determined by
the spacetime metric and which is arbitrary. This is interesting as a
toy model for the CSS solutions (where we are interested in massless
particles anyway).

In a spherically symmetric static solution in particular, the
trajectory of a particle is determined not by $E$ and $F$ separately,
but only by the combination $E^2/F$, which is invariant under
rescalings of the four-momentum. We rewrite the free function $h(E,F)$
as
\begin{eqnarray}
h(E,F)&=&k(y,F), \\
y&\equiv& \frac{E}{\sqrt{F}}=\frac{\alpha W}{\sqrt{F}}
=\alpha\left(\frac{w^2}{F}+\frac{1}{r^2}\right)^\frac{1}{2}.
\end{eqnarray}
With $m=0$, the limit $E_{\rm min}(r,F)$ in the integration over $E$
becomes independent of $F$, and the two integrations can be
interchanged. We find
\begin{eqnarray}
{\alpha'\over \alpha}&=&{a^2-1\over2r}+{4\pi^2 a^2\over r\alpha^2}
\times \nonumber \\
&& \qquad \int_{\alpha/r}^\infty \left(1-{\alpha^2\over r^2
y^2}\right)^{{1\over 2}}\bar k(y)2y\,dy, 
\label{alphamassless}
\\
{a'\over a}&=&-{a^2-1\over2r}+{4\pi^2 a^2\over r\alpha^2} \times 
\nonumber \\
&& \qquad \int_{\alpha/r}^\infty \left(1-{\alpha^2\over r^2
y^2}\right)^{-{1\over 2}}\bar k(y)2y\,dy, 
\label{amassless}
\end{eqnarray}
where $\bar k$ is the integral of the Vlasov function $k$ over
all particles with the same trajectory:
\begin{equation}
\bar k(y)\equiv\int_0^\infty k(y,F)F\,dF.
\end{equation}
The integration limit $y=\alpha(r)/r$ is the turning point
(perihelion) of all particles with a given $y$. Note that
$[f]=[g]=[h]=[k]=L^{-2}P^{-4}$ (for both massive and massless
particles), and $[\bar k]=L^2$.

For massless particles the Ricci scalar vanishes, and as a consequence
$a(r)$ and $\alpha(r)$ are not independent but are related by a second
order ODE. This ODE has a unique solution for either $a(r)$ given
$\alpha(r)$ and the boundary conditions $a(r)=1$, $a'(r)=0$, or the other
way around. $a(r)$ and $\alpha(r)$ are therefore related one-to one. 

Furthermore, Eqs. (\ref{alphamassless},\ref{amassless}) can be turned
into linear Fredholm equations of the second kind by a change of
variable.  This means that either of these equations can be solved for
$\bar k(y)$ for given $a(r)$ and $\alpha(r)$. The relationship between
$a(r)$ and $\alpha(r)$ on the one hand and $\bar k(y)$ on the other is
therefore one-to-one. 

\section{CSS spacetimes} 

\subsection{Spacetime geometry}

Because the particle mass $m$ introduces a scale into the field
equations, it is not clear that massive particles are
compatible with exact self-similarity. In this paper we therefore
construct CSS solutions of the massless Einstein-Vlasov
system. We defer the question if massive particles allow CSS
solutions or asymptotically CSS solutions to another paper.

We start by defining CSS and stating some general properties of CSS
spacetimes. Then we analyze the possible trajectories of massless
particles in those spacetimes. We find two nontrivial conserved
quantities for massless particles, $J$ and $\tau_0$, which are the
equivalents of $E$ and $t_0$ in the static case. In contrast to the
static case, we will find that the Vlasov distribution function $f$
can not just depend on $J$ and $F$ but must also depend on $\tau_0$ in
order to construct a regular self-similar stress-energy tensor.

\begin{figure*}[ht!]
\begin{center}
\includegraphics[width=15cm]{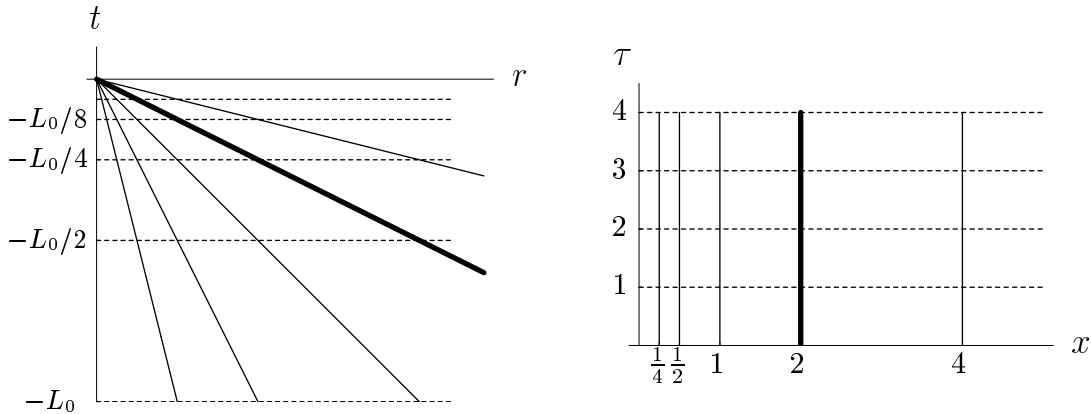}
\caption{ \label{fig:xtaucoordinates} 
Map between coordinates $(t,r)$ and $(\tau,x)$. The past light cone
of the point $t=r=0$ has been arbitrarily set to $x=x_{\rm lc}=2$.}
\end{center}
\end{figure*}

A CSS spacetime is one that possesses a homothetic vector field
\begin{equation}
\label{homothetic}
\nabla_\mu \xi_\nu + \nabla_\nu \xi_\mu = -2g_{\mu\nu} .
\end{equation}
In spherical symmetry, in the coordinates $r$ and $t$, this vector
field is
\begin{equation}
\xi=-t\partial_t-r\partial_r .
\end{equation}
$\xi$ is the infinitesimal generator of scale transformations $r\to
sr$, $t\to st$. It is customary to define the following coordinates,
which are adapted to self-similarity in the region $(r\ge 0,t<0)$:
\begin{equation}
x \equiv \frac{r}{-t} ,
\qquad
\tau \equiv -\log\left(\frac{-t}{L_0}\right) ,
\end{equation}
where $L_0$ is a constant put in to get dimensions right. It is
helpful to assign $t$, $r$, and $L_0$ dimension $L$, while $x$ and
$\tau$ are dimensionless. Fig.  \ref{fig:xtaucoordinates} gives a
graphical description of the new coordinates. In these coordinates
the homothetic vector is
\begin{equation}
\xi=\partial_\tau ,
\end{equation}
and the metric functions $a$ and $\alpha$ are functions of $x$
only. (We could write the metric explicitly in terms of coordinates
$x$ and $\tau$, but it is often more helpful to refer back to
coordinates $t$ and $r$ as they are more familiar.)

Geometrically, the self-similar spacetime singles out the point
$r=t=0$, which in fact is a curvature singularity.  It will be useful
to introduce two related auxiliary functions,
\begin{equation}
\label{Gdef}
G(x)\equiv \frac{a(x)x}{\alpha(x)}, \qquad
H(x)\equiv \frac{\alpha^2(x)}{x^2}-a^2(x).
\end{equation}
The past light cone of the singularity (from now on also referred to
as ``the'' light cone) is the set of points $(\tau, x_{\rm lc})$ with
$x_{\rm lc}$ the solution to the equation
\begin{equation}
G(x_{\rm lc})=1.
\end{equation}
(Of course $G=1$ is equivalent to $H=0$.)  In the following, we assume
that there is only a single light cone $G=1$ in the region $t<0$. 

Once we know the behavior of the metric under scale transformations
we can derive the behavior of the Einstein tensor, and therefore of
the stress-energy tensor. The components of the stress-energy tensor
in coordinates $(t,r)$ obey
\begin{equation}
{T_\mu}^\nu(t,r)=t^{-2}\bar{T}_\mu{}^\nu(x).
\label{CSSTmunu}
\end{equation}
Using the rescaled components $\bar{T}_\mu{}^\nu$, the Einstein
equations become ODEs in $x$:
\begin{eqnarray}
\frac{a'}{a}+\frac{a^2-1}{2x}
&=&
-4\pi x a^2\bar{T}_t{}^t(x) , 
\label{aprime} \\
\frac{\alpha'}{\alpha} +\frac{1-a^2}{2x}
&=&
4\pi x a^2 \bar{T}_r{}^r(x) , 
\label{alphaprime} \\
\frac{a'}{a}
&=&
4\pi a^2 \bar{T}_t{}^r(x).
\end{eqnarray}
Because in CSS $a$ is a function of $x$ only, the equations for
$a_{,t}$ and $a_{,r}$ have both become equations for $a'$, and can be
combined to eliminate $a'$ and obtain an algebraic equation for $a$,
\begin{equation}
a^{-2}(x)-1=8\pi\left[x^2\bar{T}_t{}^t(x)  +x \bar{T}_t{}^r(x)\right]
\label{aalgebraic}
\end{equation}

In a CSS spacetime, the Hawking mass $M(t,r)$ is of the form
$M(t,r)=e^{-\tau}\bar M(x)$. At the regular center, $\bar M(0)=0$, and
from the Einstein equations it follows that $\bar M'(x)\ge0$.  Suppose
now that a spherically symmetric CSS spacetime contained a vacuum
region, so that no matter was present in some interval of $x$. The
only spherically symmetric vacuum solution is Schwarzschild. But a
Schwarzschild solution has a finite Hawking mass $M(t,r)=M$, which is
incompatible with $M(t,r)=e^{-\tau}\bar M(x)$, unless $M(t,r)=0$. This
means that a spherically symmetric regular CSS solution can contain a
central vacuum region, but no vacuum region surrounding a matter
region. In particular, it can not be asymptotically flat. (The
exception would be a spacetime with a negative mass naked singularity
at the center, as $\bar M(0)<0$ would make $\bar M(\infty)=0$
compatible with $\bar M'(x)\ge 0$.)

We note in passing that for massless particles the energy-momentum
tensor is traceless, and therefore both the Einstein and Ricci tensors
are traceless. This gives an additional relation between the metric
functions $a$ and $\alpha$:
\begin{equation}
2\frac{a}{\alpha}\left(\frac{\alpha}{a}\right)'
-\frac{a^2-1}{x}
+\frac{xa}{\alpha}\left(\frac{\alpha'}{a}-x^2\frac{a'}{\alpha}\right)'
=0.
\end{equation}
Thus, we only have to solve for one of the metric functions, or
equivalently, we only have to know one of the components of the
energy-momentum tensor.

\subsection{Massless and massive particles}

We now discuss test particle trajectories in a fixed CSS spacetime in
some detail, as their properties are unfamiliar. Starting from the
homothetic vector (\ref{homothetic}) is natural to construct a
``homothetic energy'',
\begin{equation}
\label{Jdef}
J\equiv -\xi^\mu p_\mu = -p_\tau = t p_t + r p_r .
\end{equation}
In a self-similar spacetime it obeys the equation of motion
\begin{equation}
\frac{dJ}{d\sigma}=-m^2 , 
\end{equation}
where $\sigma$ is the affine parameter along particle
trajectories. For massless particles $J$ is a constant of motion. For
massive particles, $J+m^2\sigma$ is constant, but to use this constant
of motion one would have to solve for $\sigma$ as a function of
$(t,r,w,F)$.  

In a CSS background, it is natural to rewrite the Vlasov equation in
terms of $f(t,r,J,F)$, obtaining
\begin{equation}
  p^t \frac{\partial f}{\partial t}
+ p^r \frac{\partial f}{\partial r}
- m^2 \frac{\partial f}{\partial J}
= 0 ,
\label{VlasovtrJF}
\end{equation}
where $p^r$ and $p^t$ are now functions of $(t,r,J,F)$.

In order to integrate trajectories of the reduced system in $r$ and
$t$, we define
\begin{equation}
\label{vdef}
{dr\over dt}={p^r\over p^t}\equiv v .
\end{equation}
$p^r$ and $p^t$ can be obtained from $F$ and $J$ by solving the system
\begin{eqnarray}
J &=& -\alpha^2 t p^t + a^2 r p^r \label{system1}, \\
F &=& r^2 \left(\alpha^2 {p^t}^2 - a^2 {p^r}^2 - m^2 \right) 
\label{system2} .
\end{eqnarray}
We know that $F\ge 0$, and by convention $p^t>0$.  The radial momentum
$p_r$ can be positive, negative or zero, and therefore $J$ may also
have either sign.  For $a(x)x > \alpha(x)$, that is outside the light
cone, a unique solution exists for all $J$. For $a(x)x < \alpha(x)$,
that is inside the light cone, one solution exists for $J=J_c$, and
two solutions exist for $J>J_c$, where
\begin{equation}
J_c(x,F,mr) \equiv 
\sqrt{H(x)(F+m^2r^2)}.
\end{equation}
The condition $J\ge J_c$ means that we are outside or at the minimum
radius that a particle with given $x$ and $F$ can have at a given $t$.
On the light cone, $a(x)x=\alpha(x)$, we have a unique solution for
$J>J_c=0$. On the light cone $J=0$ is only possible with $F=m=0$ and 
applies to massless particles of arbitrary momentum moving radially 
into the singularity along its past light cone.

\subsection{Massless particle trajectories}

We now discuss the trajectories of massless particles in more
detail. $J$ is then a constant of the motion. From the definitions of
$v$, $J$ and $F$ it is clear that for $m=0$, $v$ depends only on $x$
and the combination
\begin{equation}
\label{Ydef}
Y \equiv {J\over \sqrt{F}}.
\end{equation}
With the convention $p^t>0$ we find that
\begin{equation}
Y=\frac{\alpha^2/x + a^2 v}{\sqrt{\alpha^2-a^2 v^2}} .
\label{vequation}
\end{equation}
To solve this equation for $v(x,Y)$, we need to square it first. This
gives rise to a sign ambiguity that we discuss below. The result is
\begin{equation}
v=\left(\frac{\alpha}{ax}\right)
\frac{-a\alpha \pm Y\sqrt{-\alpha^2+(a^2+Y^2)x^2}}{a^2+Y^2}\equiv v_\pm(x,Y).
\label{v}
\end{equation}
The allowed range of $v$ for particles with $F>0$ is
$-\alpha/a<v<\alpha/a$. The rates of change of $x$ and $r$ are related
by
\begin{equation}
\label{dxdtau}
\frac{dx}{d\tau}\equiv V=v+x
\label{velocities}
\end{equation}
In coordinates $x$ and $\tau$, the particle equation of motion is
autonomous: 
\begin{equation}
{dx\over d\tau}=v_\pm(x,Y)+x\equiv V_\pm(x,Y).
\end{equation}
Trajectories of massless particles are therefore determined by $Y$, up
to a translation in $\tau$. A translation in $\tau$ corresponds to a
simultaneous rescaling of $t$ and $r$. A second symmetry of massless
particle trajectories arises because $Y$ is invariant under a
simultaneous rescaling of all components of $p^\mu$. This leaves the
trajectory in spacetime completely invariant: photons of different
energy can have the same trajectory. The overall momentum scale, at
constant $Y$ and constant $r$ and $t$, is set by $\sqrt{F}$. 
Here we assume that the contribution of particles with $F=0$ to the 
stress-energy is zero, and we now concentrate on particles with $F>0$.
(If they made a finite contribution, this would constitute null dust.
In our numerical examples we avoid this simply by imposing a cutoff.)

\begin{figure}[t]
\begin{center}
\includegraphics[width=8cm]{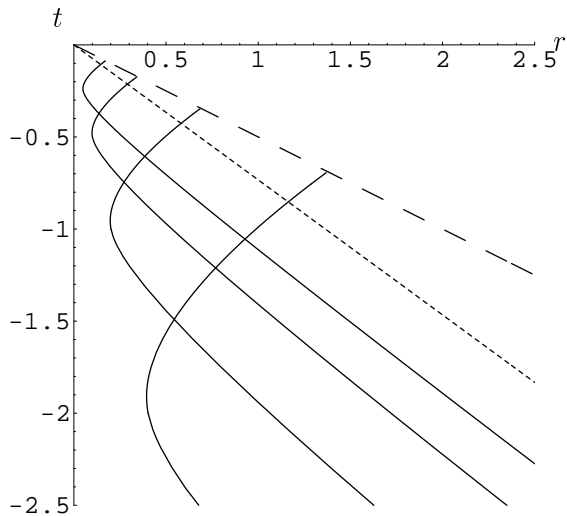}
\caption{ \label{fig:sameY} Trajectories of particles in the
self-similar spacetime described in Sect. \ref{CSSsolution} below, all
with the same value $Y=5$ but different overall scales. The past light
cone of the singularity at $x_{\rm lc}=1.3642$ is denoted with a 
short-dashed line, and for orientation $x=2$ has been marked with a 
long-dashed line.}
\end{center}
\end{figure}

\begin{figure}[t]
\begin{center}
\includegraphics[width=8cm]{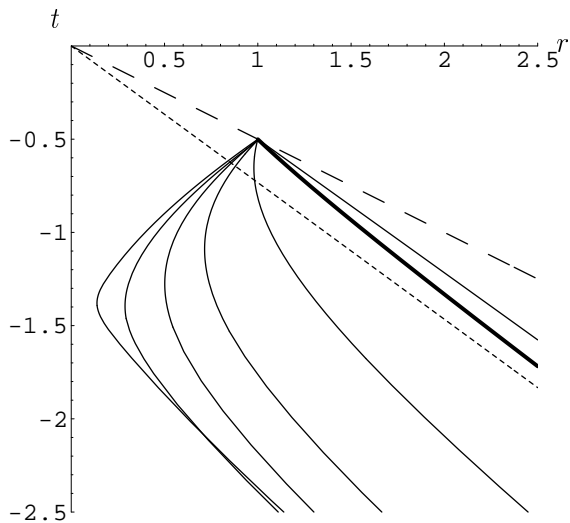}
\caption{ \label{fig:rt} $(t,r)$-trajectories of particles in the
self-similar spacetime \ref{CSSsolution} with $Y=10,5,3,2,1,0,-1$
(from left to right).
The trajectory with $Y=0$ has been indicated by a thicker line.  The
overall scale of each trajectory has fixed so that all
trajectories meet in one point.}
\end{center}
\end{figure}

\begin{figure}[t]
\begin{center}
\includegraphics[width=8cm]{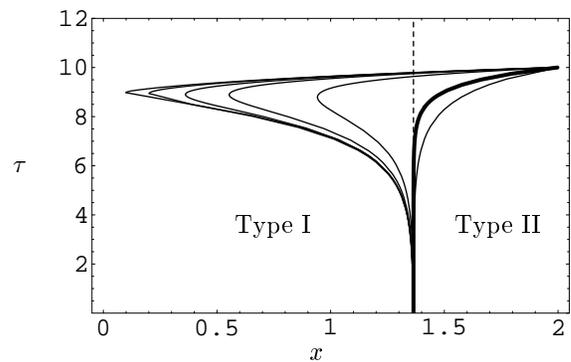}
\caption{ \label{fig:xtau} The same trajectories in coordinates
$(x,\tau)$. From left to right we have $Y=10,5,3,2,1,0,-1$. The
thicker line is $Y=0$. The origin in the $\tau$ axis has been
arbitrarily chosen (corresponding to a choice of units $L_0$).  }
\end{center}
\end{figure}

\begin{figure}[t]
\begin{center}
\includegraphics[width=8cm]{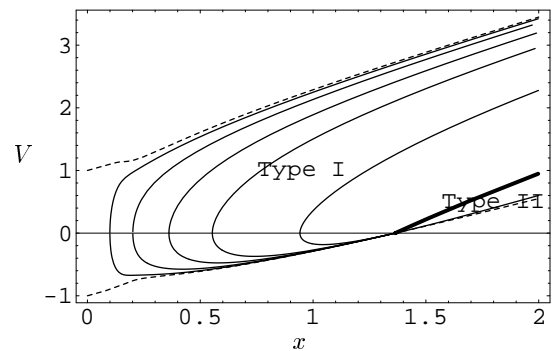}
\caption{ \label{fig:xV}
The equivalent trajectories (from left to right $Y=10,5,3,2,1,0,-1$) 
in variables $x$ and $V$. 
}
\end{center}
\end{figure}

The projection into the $(r,t)$ plane of a the trajectory of a
massless particle with $F>0$ is timelike. Such particles come in from
infinity, reach a minimum radius $r$, and go out to infinity again. At
large $r$, the tangential velocity can be neglected compared to the
radial velocity, and the radial projection of the trajectory is
asymptotically null. Fig. \ref{fig:sameY} shows several trajectories
in the $(r,t)$ plane that share the same value of $Y$ but are related
by a rescaling of $r$ and $t$. In the $(x,\tau)$ plane these would be
related by a translation in $\tau$. Fig. \ref{fig:rt} shows typical
particle trajectories in the $(r,t)$ plane with different values of
$Y$. Fig. \ref{fig:xtau} shows the same trajectories in $(x,\tau)$
coordinates, and Fig. \ref{fig:xV} shows the equivalent trajectories
in $(x,V)$ space. One should keep in mind that the turning point of
smallest $x$ (if one exists) does not coincide with the turning point
of smallest $r$.

Fig. \ref{fig:xtau} illustrates that there are two types of particles:
type I particles are initially inside the light cone but leave it,
while type II particles are always outside the light cone. Both types
of particles start at $x=x_{\rm lc}$ in the asymptotic past as
$\tau\to-\infty$. Type I particles peel off from $x=x_{\rm lc}$
towards smaller $x$, reach a minimum value $x_{\rm min}(Y)$ of $x$ and
turn round. A finite $\tau$-time later they cross the light cone
$x=x_{\rm lc}$, and continue to large $x$. Type II particles peel off
from $x=x_{\rm lc}$ towards larger $x$. Although they reach a minimum
value of $r$, their value of $x$ always increase. Both types of
particles reach $x=\infty$ at $\tau=\infty$. This is just the surface
$t=0$, where our coordinate system breaks down. Here we are only
interested in the region $t<0$.

In order to see when $dx/d\tau$ vanishes, and which of the two values $V_\pm$
is the correct one in a given situation, it is helpful to factorize
$V_\pm$ as
\begin{equation}
\label{Vfactors}
V_\pm(x,Y)=x(a^2+Y^2)^{-1}(Y^2-H)^{1/2}\left[(Y^2-H)^{1/2}\pm
G^{-1}Y\right],
\end{equation}
where $G(x)$ and $H(x)$ were defined in (\ref{Gdef}). Clearly both
$V_+$ and $V_-$ vanish at $x=x_{\rm min}(Y)$ which is defined by
\begin{equation}
H(x_{\rm min})\equiv Y^2,
\label{turningpoints}
\end{equation}
One can show that at these points $d^2x/d\tau^2$ does not vanish and
has the same sign as $-dH/dx$. Therefore these points are true turning
points. The square bracket in (\ref{Vfactors}) can vanish only at the
light cone $x=x_{\rm lc}$, where $H=0$ and $G=1$. In particular $V_+$
vanishes at $x=x_{\rm lc}$ for all $Y<0$ (but not for $Y>0$), and
$V_-$ vanishes for all $Y>0$ (but not for $Y<0$). This corresponds to
$V\to 0$ as $\tau\to-\infty$ (compare Fig. \ref{fig:xtau}), and one
can show that $d^2x/d\tau^2=0$ there.

To see how the sign choice $V_\pm$ relates to our classification of
particles, we note first that $Y$ and $-Y$ give rise to inequivalent
trajectories. With $t<0$ and $p^t>0$, it follows from (\ref{system1})
that $J>0$ for any particle inside the light cone. Therefore type I
particles have $Y>0$. Inside the light cone, the velocity of type I
particles is $V=V_-<0$ while they are ingoing, and $V=V_+>0$ while
they are outgoing. By continuity, $V=V_+$ also outside the light cone.

From Fig. \ref{fig:xV} we see that type II particles have $v<0$, and
therefore $p^r<0$. (To be precise, this holds for the region $t<0$
covered by our coordinates. Presumably these particles have an
$r$-turning point for some $t>0$. $Y=0$ particles have an $r$-turning
point at $t=0$.) Therefore type II particles have $J<0$, and so
$Y<0$. This is summarized in Table \ref{table:V}. 

\begin{table}
\caption{ \label{table:V} 
Type I and type II particles
}
\begin{ruledtabular}
\begin{tabular}{lll}
& Type I & Type II \\ 
& $Y>0$ & $Y<0$ \\
inside light cone &  $V=V_\pm$ & never there \\
outside light cone & $V=V_+$ & $V=V_+$ 
\end{tabular}
\end{ruledtabular}
\end{table}

It seems plausible that CSS spacetimes exist where $H(x)$ has a local
minimum, so that for some range of $Y^2$ the turning point equation
$H=Y^2$ would have two solutions $x_{\rm min}$ and $x_{\rm max}$ for
each value of $Y$. Particles with $Y$ in that range would oscillate
between the two turning points. Such ``type III'' particles would be
trapped within a finite range of $x$, and would therefore hit the
singularity. However, from the definition (\ref{Jdef}) of $J$ it
follows that $p^r$ and $p^t$ at constant $x$ and $J$ scale as
$t^{-1}$. As a type III particle returns to the same value of $x$
again and again, its energy and radial momentum as measured by a
constant $r$ observer increase as $t^{-1}$. A similar argument holds
for the tangential momenta. A finite number of such particles would
therefore constitute an energy-momentum scaling as $t^{-1}$ confined
to a region of a physical size scaling as $t^{-1}$, so that the energy
and other frame components of the stress-energy would scale as
$t^{-4}$. This is incompatible with self-similarity, where these
stress-energy components scales as $t^{-2}$. Type III particles are
allowed as test particles, but can not be used as the matter content of
a CSS spacetime. From now on we restrict consideration to type I and
II particles.

\subsection{Solutions with massless particles}
\label{CSSt0}

For massless particles, $J$ is a constant of motion, and therefore
$\partial/\partial J$ disappears from the Vlasov equation
(\ref{VlasovtrJF}). We then obtain an automatic solution of the Vlasov
equation by the ansatz $f(t,r,J,F)=h(J,F)$. However, we shall see that
this ansatz leads to a divergent stress-energy tensor. Therefore we
have to resort to the most general situation on a CSS background, by
allowing for the dependence of $f$ on the spacetime coordinates $r$
and $t$ through a dependence on a third non-trivial constant of the
motion. (With $c_1=J$, this is the constant $c_2$ of our general
discussion of conserved quantities.) We rewrite the Vlasov equation
for massless particles in coordinates $x$ and $\tau$ as
\begin{equation}
\left(\frac{\partial}{\partial\tau}+V(x,Y)\frac{\partial }{\partial
x}\right)f(\tau,x,J,F)=0,
\end{equation}
We can solve this by the method of characteristics as
\begin{equation}
f(\tau, x, J, F) =
g\left(\tau_0,J,F\right),
\end{equation}
where \begin{equation} \tau_0(\tau,x,Y) \equiv \tau -
\int_{x_0(Y)}^{x}\frac{dx'}{V(x',Y)} .
\label{tau0constant}
\end{equation}
Clearly $\tau_0$ is the time $\tau$ when the particle with a given $Y$
and $F$ was at $x=x_0$. Here we have specified $x_0$ as a given
function of $Y$ (which could be a constant). A priori $x_0$ could also
depend on $F$, but that would break self-similarity of the resulting
stress-energy.  For $x_0(Y)<x_{\rm lc}$ (only possible for type I
particles) we must specify if the spacetime point $(x_0,\tau_0)$ is on
the ingoing or outgoing part of the trajectory. 
A natural choice then is $x_0(Y)=x_{\rm
min}(Y)$, but this does not work for type II particles, which do not
have a turning point. For either type of particle we could use a fixed
value $x_0={\rm const}>x_{\rm lc}$. 
As we are mainly working in coordinates $t$ and $r$ in this paper, we 
shall use $\tau_0$ in the form of
\begin{equation}
t_0\equiv -L_0 e^{-\tau_0}=t\exp\int_{x_0(Y)}^x \frac{dx'}{V(x',Y)}\equiv
t\,Q(x,Y).
\end{equation}
(Note again that $Q(x,Y)$ depends on the choice of $x_0(Y)$.)

We have just shown that the general spherically symmetric solution of
the Vlasov equation in a spherically symmetric CSS spacetime is
\begin{equation}
\hbox{Vlasov in CSS spacetime} \quad \Rightarrow \quad
f(t,r,w,F)=g\left(t_0,J,F\right) .
\end{equation}
One could think, by analogy with the static case, that the requirement
of a self-similar energy-momentum tensor eliminates the dependence on
$t_0$. However, that is not the case. In the static case, the
constants of motion $E$ and $F$ depend only on the variables $r$,
$p^t$ and $p^r$. Only the constant of motion $t_0$ depends also on
$t$, and that is why $g$ can not depend on it (except in the piecewise
constant way of Eq. (\ref{GRJeans})). In the self-similar case, not
just $t_0$ but all three constants $t_0$, $J$ and $F$ change under a
scale transformation. Imposing consistency with a CSS spacetime is
therefore less straightforward.

In order to derive the appropriate form of $g(t_0,J,F)$, it is helpful
to change the integration over $F$, which has dimension $P^2L^2$,
temporarily to an integration over $\bar F=F/r^2$, which has dimension
$P^2$. This completely separates the integration over momentum space
from the remaining dependence on $r$ and $t$. The rescaled
stress-energy tensor components are
\begin{eqnarray}
\bar{T}_r{}^r&=&\pi \, \int_0^\infty d\bar F
\int_{-\infty}^{\infty} dw \, t^2 g(t_0,J,F)
\frac{w^2}{W} 
\\
\bar{T}_t{}^t&=&-\pi \, \int_0^{\infty} d\bar F
\int_{-\infty}^{\infty} dw \, t^2 g(t_0,J,F) W ,
\\
\bar{T}_t{}^r&=&-\pi \frac{\alpha}{a} \, \int_0^{\infty} d\bar F
\int_{-\infty}^{\infty} dw \, t^2 g(t_0,J,F) w ,
\\
\bar{T_\theta}^\theta&=&\frac{\pi}{2} \, \int_0^{\infty} d\bar F
 \int_{-\infty}^{\infty} dw \, t^2 g(t_0,J,F) \frac{\bar F}{W}.
\end{eqnarray}
The functions $t_0$, $J$, $F$, and $W$ are homogeneous functions of
$r$ and $t$ of degree 1, 1, 2 and 0, respectively. If we now demand
that $\bar{T}_\mu{}^\nu(st,sr)=\bar{T}_\mu{}^\nu(t,r)$ for any scaling
constant $s$, so that the stress-energy tensor is compatible with CSS,
we find that $g$ must be the following homogeneous function of its
arguments:
\begin{equation}
{\rm Vlasov + CSS} \quad \Rightarrow \qquad 
g(st_0,sJ,s^2F)=s^{-2}g(t_0,J,F)
\label{Vss}
\end{equation}
A homogeneous function of three variables can be re-expressed as an
arbitrary function of two variables. We define
\begin{equation}
g(t_0,J,F)=\frac{1}{F}k(Y,Z),\quad \hbox{ where } Z\equiv
\frac{F}{t_0^2}
\label{finalh}
\end{equation}
and where $Y$ was defined above in (\ref{Ydef}).  The situation is
similar to static solutions of Einstein-Vlasov in that we can freely
specify a function $k(Y,Z)$ of two variables that automatically solves
the Vlasov equation, and obtain the spacetime by solving the Einstein
equations, which become integral-differential equations. It differs in
that $g(t_0,J,F)$ is not simply of the form $h(J,F)$. Because we need
$t_0$ in order to obtain the general solution, we need to determine
$Q(x,Y)$, which is equivalent to integrating the equations of
motion. Note that $f(t,r,w,F)$ is determined by $h(Y,Z)$ only when the
function $x_0(Y)$ that appears in $Q$ has been specified.

It is interesting to note that the homogeneity condition (\ref{Vss})
arises from dimensional analysis in the length dimension $L$
alone. $Y$ has neither $P$ nor $L$ dimension, but $Z$ has dimension
$P^2$. This is not an obstruction to self-similarity of the
spacetime. If we had measured particle energy-momentum in units of
length, dimensional analysis would have confused us in this point.
Note that $[f]=[g]=L^{-2}P^{-4}$, $[k]=P^{-2}$, and $[\bar k]=L^2$. 

The Einstein equations are now once again a set of
integral-differential equations that we can solve by iteration, once
$k(Y,Z)$ and $x_0(Y)$ have been specified. The next step in our
program would be to change the integration variables from $F$ and $w$
to $Y$ and $Z$. By analogy with the massless static case, we expect
the integration over $Z$ to decouple from that over $Y$. This is
correct, but going to the variables $Y$ and $Z$ directly gives rise to
rather complicated expressions. When calculating the stress-energy
components, we find it helpful to integrate over the variables $Z$ and
$u\equiv w/\sqrt{\bar F}$. This already decouples the integrals:
\begin{eqnarray}
\bar{T}_r{}^r &=&\frac{\pi}{x^4}\int_{-\infty}^\infty 
Q^2(x,Y) \, \bar{k}(Y) \frac{u^2}{\sqrt{u^2+1}} \, du , \\
\bar{T}_t{}^t &=&-\frac{\pi}{x^4}\int_{-\infty}^\infty 
Q^2(x,Y) \, \bar{k}(Y) \sqrt{u^2+1} \, du , \\
\bar{T}_t{}^r &=&-\frac{\pi}{x^4}\frac{\alpha}{a}\int_{-\infty}^\infty 
Q^2(x,Y) \, \bar{k}(Y) u \, du , \\
\bar{T}_\theta{}^\theta &=& \frac{\pi}{2x^4}\int_{-\infty}^\infty 
Q^2(x,Y) \, \bar{k}(Y) \frac{1}{\sqrt{u^2+1}} \, du , 
\end{eqnarray}
where
\begin{equation}
\bar k(Y)\equiv\int_0^\infty k(Y,Z)\,dZ.
\label{barkdef}
\end{equation}
The auxiliary variable $u$ is related to $Y$ by
\begin{equation}
Y=Y(x,u)=\frac{\alpha(x)}{x}\sqrt{u^2+1}+a(x)u.
\end{equation}
We can now see clearly why $g$ must depend on $t_0$. If it did not,
$k(Y,Z)$ could not depend on $Z$, and therefore $\bar k(Y)$ would
diverge. The dependence on $t_0$ is necessary to create a finite CSS
stress-energy tensor.

\subsection{Some explicit solutions}
\label{CSSsolution}

We begin with a CSS solution of the Vlasov system in flat spacetime,
which is trivially self-similar with $x_{\rm lc}=1$. $Q_\pm(x,Y)$ can
then be calculated in closed form. 
From the definition of $Q$ it follows that it obeys the differential 
equation
\begin{equation}
{\partial\ln Q_\pm(x,Y)\over\partial x}={1\over V_\pm(x,Y)}. 
\end{equation}
Integration of this is straightforward when $a(x)=\alpha(x)=1$.  We
now consider the boundary conditions. 
We make the assumption that the reference point $(x_0,\tau_0)$ is on the
outgoing branch of the trajectories. This gives rise to the boundary 
condition $Q_+(x_0(Y),Y)=1$ for outgoing particles. For ingoing 
particles we have to piece $Q$ together from an ingoing and an outgoing
piece, integrating first from $x$ to $x_{\rm min}$ using $V_-$ and then
from $x_{\rm min}$ to $x_0$ using $V_+$, respectively.
Here the turning point $x_{\rm min}$ of type I particles is given by
\begin{equation}
x_{\rm min}(Y)=(1+Y^2)^{-1/2}
\end{equation}
for $Y>0$. (Type II particles, with $Y<0$, have no turning point.)
Our final result is 
\begin{equation}
\label{flatQ}
Q_\pm(x,Y)=\frac{Y\pm\sqrt{x^2(1+Y^2)-1}}{Y+\sqrt{x_0^2(Y)(1+Y^2)-1}},
\end{equation}
which holds for type I and type II particles as long as $x$ and
$x_0(Y)$ are both in the physical range: $x>x_{\rm min}(Y)$ for $Y>0$,
and $x>1$ for $Y<0$. Note that $Q_-\to 0$ as $x\to 1_-$ for type I
particles and $Q_+\to 0$ as $x\to 1_+$ for type II particles: this is 
the limit in
which the trajectory peels off the light cone at $\tau\to-\infty$. Note
also that $Q_\pm$ is finite for type I particles as $x_0\to 1$ because
the reference point is the point where the trajectory crosses the
light cone while going out. By contrast, $Q_+$ diverges for type II
particles as $x_0\to 1_+$ because this pushes the reference point to
$\tau\to-\infty$ where the trajectory peels off the light cone.

On a flat spacetime, neglecting gravity, we are just dealing with the
Vlasov equation, which is linear. It is therefore sufficient to
consider the following ansatz for $\bar k(Y)$:
\begin{equation}
\bar{k}(Y)=\bar{k}_+ \, \delta(Y-Y_0) + \bar{k}_- \, \delta(Y+Y_0),
\end{equation}
with $Y_0>0$. We can then integrate the resulting stress-energy tensor
over $Y_0$, with arbitrary functions $\bar k_\pm(Y_0)$, in order to
obtain the general result. We shall see later why it is useful to
consider the contributions $Y=Y_0$ and $Y=-Y_0$ together.

We first calculate the stress-energy inside the light cone. Because
type II particles are always outside the light cone, a solution with
both type I and type II particles is independent of its type II
particle content for $0\le x<1$. Inside the light cone we find that
the stress-energy tensor and particle number current components are
\begin{eqnarray}
\bar{T}_t{}^t (x) &=& 
  -\frac{2\pi K_+}{x_{\rm min}^3} \, 
   \frac{1}{x\sqrt{x^2-x_{\rm min}^2}} ,\\
\bar{T}_r{}^r (x) &=&  
   \frac{2\pi K_+}{x_{\rm min}^3} \, 
   \frac{x_{\rm min}^4+(1-x_{\rm min}^2)(x^2-x_{\rm min}^2)}{x^3\sqrt{x^2-x_{\rm min}^2}} ,\\
\bar{T}_t{}^r (x) &=&  
   \frac{2\pi K_+}{x_{\rm min}} \, 
   \frac{1}{x^2\sqrt{x^2-x_{\rm min}^2}} ,\\
\bar{T}_\theta{}^\theta (x) &=&  
   \frac{\pi K_+}{x_{\rm min}} \, 
   \frac{1+x^2-2x_{\rm min}^2}{x^3\sqrt{x^2-x_{\rm min}^2}} ,\\
\bar{N}^t (x) &=& 
   \frac{2\pi K_+}{x_{\rm min}^2} \, 
   \frac{\sqrt{1-x_{\rm min}^2}}{x^2\sqrt{x^2-x_{\rm min}^2}} ,\\
\bar{N}^r (x) &=& 
   \frac{2\pi K_+}{x_{\rm min}^2} \, \frac{\sqrt{1-x_{\rm min}^2}}{\sqrt{x^2-x_{\rm min}^2}}
   \frac{x^2-2x_{\rm min}^2}{x^3}.
\end{eqnarray}
for $x>x_{\rm min}$, and vanish for $x<x_{\rm min}$.  Here $x_{\rm
min}$ is shorthand for $x_{\rm min}(Y_0)$, and we have introduced the
shorthands
\begin{equation}
K_\pm\equiv \frac{\bar{k}_\pm}{\left(Y_0 \pm
\sqrt{x_0^2(1+Y_0^2)-1}\right)^2} ,
\end{equation}
These expressions already take into account the contribution of both
ingoing and outgoing particles. We have used $Q$ given by
Eq. (\ref{flatQ}) so that the reference point $(x_0,\tau_0)$ is
always assumed to be on the outgoing branch. All these expressions
vanish for $x<x_{\rm min}$, where no particles reach, and
have an integrable divergence at $x=x_{\rm min}$, where particles pile
up as they turn around.

We now calculate the stress-energy outside the light cone. For type II
particles, we must use $x_0(Y)>x_{\rm lc}=1$. In the following, we use
the same value $x_0>1$ for both $x_0(Y_0)$ and $x_0(-Y_0)$. It is
therefore automatically on the outgoing branch of type I
particles. Taking into account the contributions of both type I and
type II particles, we find for $x>1$ that
\begin{widetext}
\begin{eqnarray}
\bar{T}_t{}^t (x) &=& 
  -\frac{\pi}{x_{\rm min}^3} \, 
   \frac{K_++K_-}{x\sqrt{x^2-x_{\rm min}^2}} , \\
\bar{T}_r{}^r (x) &=&  
   \frac{\pi}{x_{\rm min}^3} \, \left[
   (K_++K_-)\frac{x_{\rm min}^4+(1-x_{\rm min}^2)(x^2-x_{\rm min}^2)}{x^3\sqrt{x^2-x_{\rm min}^2}} 
   -2(K_+-K_-)\frac{x_{\rm min}^2\sqrt{1-x_{\rm min}^2}}{x^3}
   \right] , \\
\bar{T}_t{}^r (x) &=&  
   \frac{\pi}{x_{\rm min}} \, \left[
   (K_++K_-)\frac{1}{x^2\sqrt{x^2-x_{\rm min}^2}} 
   -(K_+-K_-)\frac{\sqrt{1-x_{\rm min}^2}}{x^2x_{\rm min}^2}
   \right] , \\
\bar{T}_\theta{}^\theta (x) &=&  
   \frac{\pi}{2x_{\rm min}} \, \left[
   (K_++K_-)\frac{1+x^2-2x_{\rm min}^2}{x^3\sqrt{x^2-x_{\rm min}^2}} 
   +2(K_+-K_-)\frac{\sqrt{1-x_{\rm min}^2}}{x^3}
   \right] , \\
\bar{N}^t (x) &=& 
   \frac{\pi}{x_{\rm min}^2} \,  \left[
   (K_++K_-)\frac{1}{x^2} 
   +(K_+-K_-)\frac{\sqrt{1-x_{\rm min}^2}}{x^2\sqrt{x^2-x_{\rm min}^2}}
   \right] , \\
\bar{N}^r (x) &=& 
   \frac{\pi}{x_{\rm min}^2} \, \left[
   (K_++K_-)\frac{1-2x_{\rm min}^2}{x^3}
   +(K_+-K_-)\frac{\sqrt{1-x_{\rm min}^2}}{\sqrt{x^2-x_{\rm min}^2}}
   \frac{x^2-2x_{\rm min}^2}{x^3}
   \right] .
\end{eqnarray}
\end{widetext}
The constant $K_-$ was defined above. Here, as above, we use $x_{\rm
min}$ as shorthand for the turning point $x_{\rm min}(Y_0)$ of the
type I particles. (This number also appears in the contribution of the
type II particles, but is then just used as a shorthand.)

Based on these results, we can make two important general points. The
first is that for general values of $K_\pm$ the stress-energy tensor
is not continuous at the light cone. However, if and only if $K_+=K_-$
the stress-energy is continuous at the light cone, and in fact
analytic. On the other hand, the particle-current is always
continuous, but never analytic. In the general flat-space case, where
$\bar k_\pm$ are promoted to functions of $Y>0$, the stress-energy is
analytic if $\bar K_+(Y)=\bar K_-(Y)$ for all $Y$. 

The second important point is that while we can freely specify both $
k(Y,Z)$ and $x_0(Y)$, the location of the reference point $x_0(Y)$ is
not really physical. We have seen that the stress-energy in our
example does not depend on the two numbers $x_0$ and $\bar k_+$ (or
$\bar k_-$) separately but only on the combination $\bar K_+$ (or
$\bar K_-$). Generally, the Vlasov density $f(t,r,w,F)$ is completely
determined by one free function of $Y$ and $Z$. In flat spacetime, and
assuming that $x_0(Y)=x_0(|Y|)$, we have shown that
this free function is 
\begin{equation}
K(Y,Z)\equiv \frac{k(Y,Z)}{\left(|Y| +
\sqrt{x_0^2(Y)(1+Y^2)-1}\right)^2} .
\end{equation}
In the self-gravitating case we cannot give this function in closed
form. Therefore, in the self-gravitating case we must just ``fix the
gauge'' by fixing $x_0(Y)$ arbitrarily.

\begin{figure}[t]
\begin{center}
\includegraphics[width=8cm]{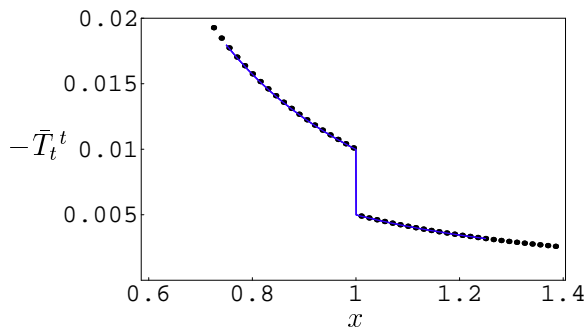}
\caption{ \label{fig:rhojump}
Discontinuity in $-\bar{T}_t{}^t$ for the $\delta$-function case with 
only type I particles (continuous line). The dots denote the 
same quantity for the Gaussian function (\ref{Gaussian}) with the same
$\bar{k}_+/\sqrt{2\pi}=10^{-4}$, $Y_0=6$ and $\sigma=1$. The agreement 
is very good.}
\end{center}
\end{figure}

We shall now smear the $\delta$-function in $Y$, and then couple this
distribution to gravity. In the following we restrict our ansatz to
type I particles. Inside the light cone this is no restriction
anyway. It allows us to use $x_0(Y)=x_{\rm min}(Y)$, which simplifies
the numerical calculations.  We also want to start with a matter
distribution whose stress-energy tensor in flat spacetime is
discontinuous, because we shall see that the gravitational
back-reaction makes it continuous. As an example, the jump in
$-\bar{T}_t{}^t$ is shown in Fig. \ref{fig:rhojump} for the case 
$Y_0=6$, $\bar{k}_+=10^{-4}\sqrt{2\pi}$ and $\bar{k}_-=0$, using 
$x_0(Y)=x_{\rm min}(Y)$.

\begin{figure}[t]
\begin{center}
\includegraphics[width=8cm]{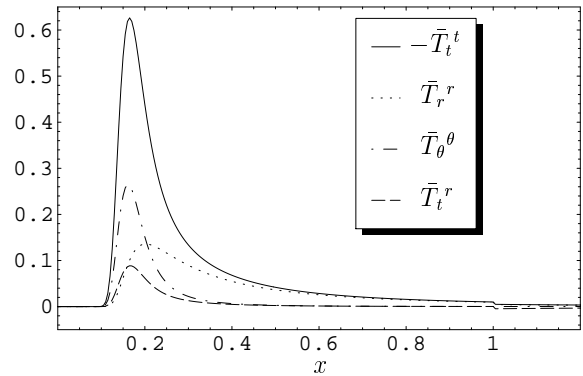}
\caption{ \label{fig:flatmatter}
Stress-energy tensor corresponding to Gaussian (\ref{Gaussian})
on flat spacetime. Note the jump at $x=x_{\rm lc}=1$.
}
\end{center}
\end{figure}

Still in flat spacetime, we smear the $\delta$-function into a Gaussian:
\begin{equation}
\bar{k}(Y)=\frac{\bar{k}_+}{\sqrt{2\pi}\sigma} e^{-(Y-Y_0)^2/2\sigma^2} 
\label{Gaussian}
\end{equation}
with a minimum cutoff just above $Y=0$ and a maximum cutoff far from
$Y_0$. It is not possible to calculate the results analytically, so
that we calculate them with a C code. The divergence of the
stress-energy at $x_{\rm min}(Y)$ is then smoothed out by the
integration over $Y$, and the stress-energy components are very smooth
apart from at the light cone. The results for the stress-energy tensor
are shown in Fig. \ref{fig:flatmatter} for the case $\sigma=1$ with
$Y_0=6$ and $\bar{k}_+=10^{-4}\sqrt{2\pi}$ as in the $\delta$-function
example. 

\begin{figure*}[ht!]
\begin{center}
\includegraphics[width=8cm]{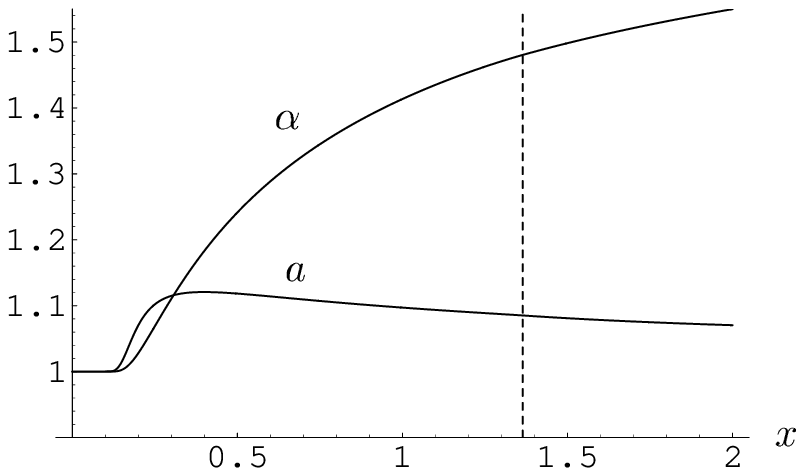}
\includegraphics[width=8cm]{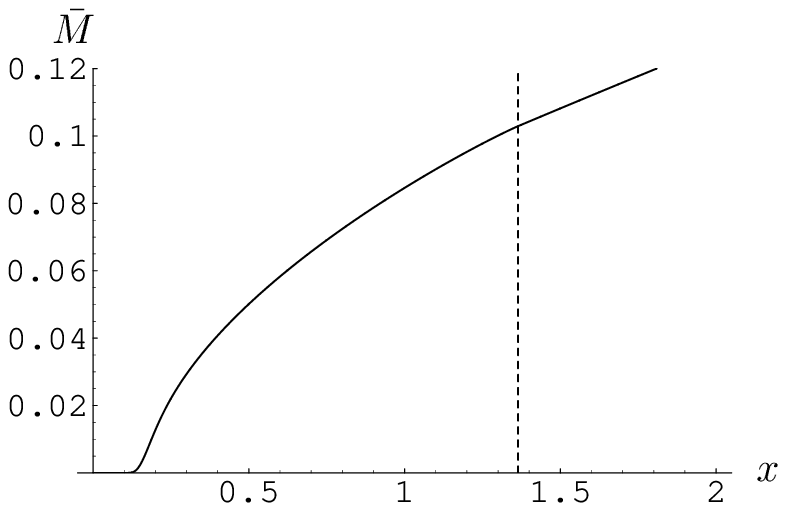}
\caption{ \label{fig:aandm}
Metric function $a(x)$ and corresponding (dimensionless) mass function 
$\bar M(x)$. Note that the spacetime is not asymptotically flat,
as we expected. The vertical lines give the position of the light cone
$x_{\rm lc}=1.3642$.
}
\end{center}
\end{figure*}

\begin{figure*}[ht!]
\begin{center}
\includegraphics[width=8cm]{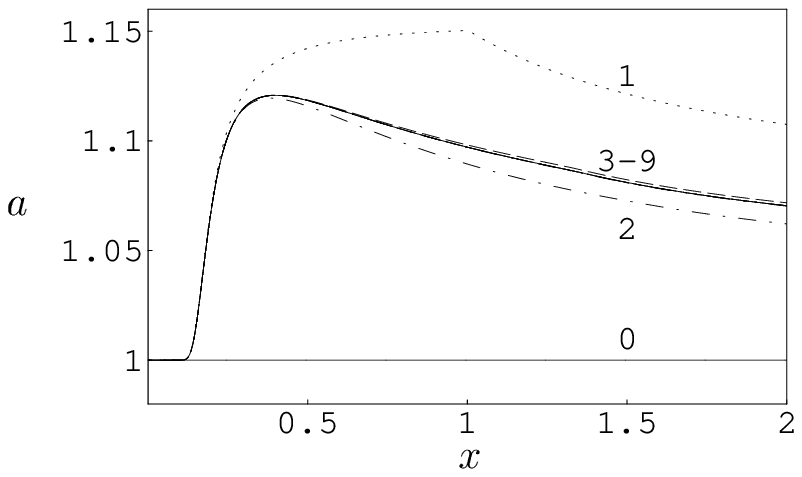}
\includegraphics[width=9cm]{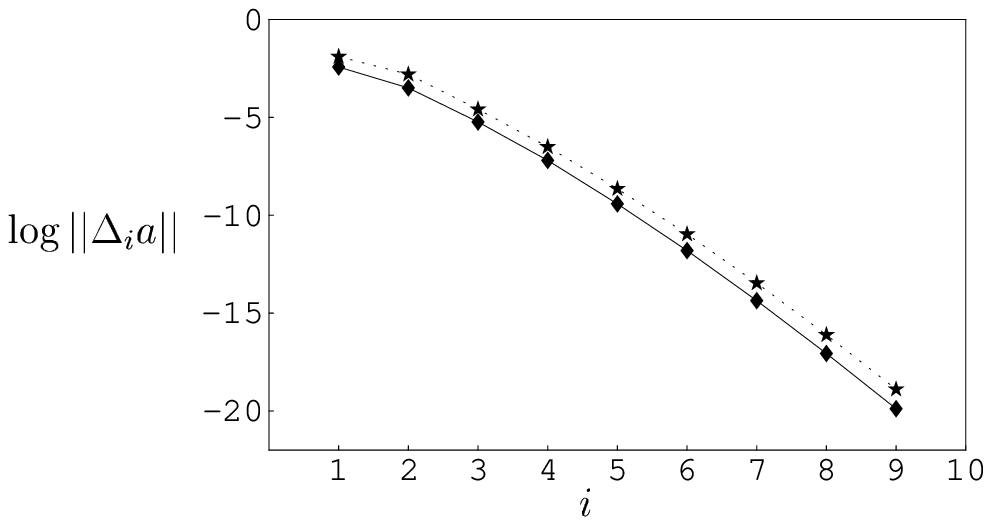}
\caption{ \label{fig:convergence}
On the left we show the different iterations of the metric function
$a(x)$, starting with the flat case in the iteration $i=0$. The
convergence is fast and starting from $i=3$ or $i=4$ it is not possible
to resolve different iterations in the figure. On the right, we show
the decay of differences between successive iterations. We have defined
$\Delta_i a \equiv a_i-a_{i-1}$. The continuous line represents the
2-norm and the dotted line the $\infty$-norm, both integrated between
$x=0$ and $x=2$.
}
\end{center}
\end{figure*}

\begin{figure}[t]
\begin{center}
\includegraphics[width=9cm]{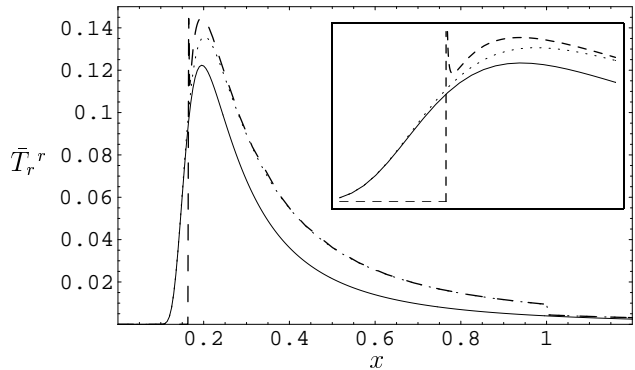}
\caption{ \label{fig:pressure}
Pressure function $\bar{T}_r{}^r(x)$. The $\delta$-function case in flat
space is represented with a dashed line; the Gaussian coupled to gravity
is given by the continuous line and the Gaussian in flat spacetime is 
given by a dotted line which interpolates between the others. The box on
the right shows an enlargement around the maximum.
}
\end{center}
\end{figure}

\begin{figure}[t]
\begin{center}
\includegraphics[width=8cm]{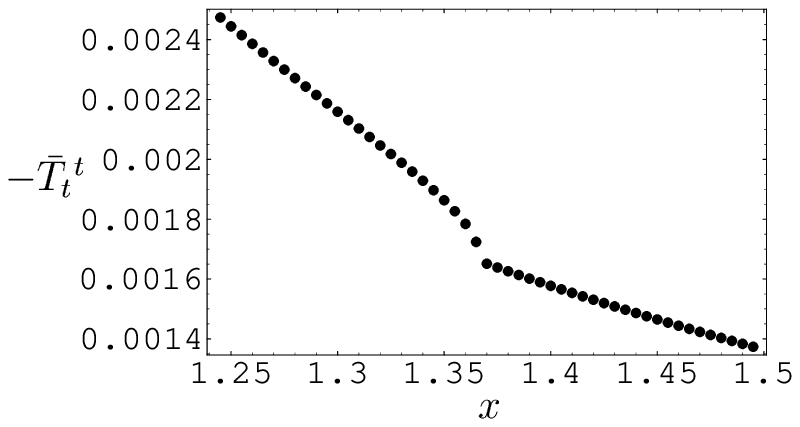}
\caption{ \label{fig:rhonojump}
When Vlasov matter is coupled to gravity, the stress-energy tensor is
continuous, but not $C^1$.
}
\end{center}
\end{figure}

Once we couple the matter to gravity we must solve the
integral-differential equations by iteration. That is, we start with
some initial metric, say flat spacetime, and calculate the
stress-energy tensor of a set of particles for a certain function
$\bar{k}(Y)$. Then we integrate the metric corresponding to this
stress-energy distribution using Eqs. (\ref{aprime}, \ref{alphaprime})
and calculate a new stress-energy tensor, and so on,
until the process converges.  In Fig. \ref{fig:aandm} we show the
metric that results from coupling the Gaussian $\bar{k}$ of the
previous example, with parameters $Y_0=6$, $\sigma=1$ and
$\bar{k}_0=10^{-4}\sqrt{2\pi}$, to gravity. The convergence of the
numerical method is demonstrated in Fig.
\ref{fig:convergence}. Fig. \ref{fig:pressure} compares the radial
pressure profile of the Gaussian ansatz in flat spacetime with the
Gaussian coupled to gravity. Even though we started with a
discontinuous stress-energy tensor on flat spacetime, we now find that
the self-consistent stress-energy is now $C^0$ (but not $C^1$). This
is illustrated in Fig. \ref{fig:rhonojump}. As a consequence, the
metric is $C^1$. Another check of the procedure is given by the Eq.
(\ref{aalgebraic}).

With very low values of $\bar{k}_+$ we can form solutions with a metric
which is very close to flat spacetime. On the other hand, with larger
values of $\bar{k}_+$ we get solutions which are very close to horizon
formation. For example with $\bar{k}_+=3\cdot 10^{-4}\sqrt{2\pi}$,
the maximum value of $a$ is greater than 1.45 (giving 
$2M/r\simeq 0.53$), which is even bigger than the maximum value of the
same function in the Choptuik or Evans-Coleman spacetimes.

We can understand analytically why the coupling to gravity makes the
matter stress-energy more regular. For particles that just peel off
the light cone, that is for $x\simeq x_{\rm lc}$, and $V=V_-$ for type
I particles and $V=V_+$ for type II particles, we can expand $V$ as
\begin{eqnarray}
V_\pm(x,\mp|Y|)
= c (x-x_{\rm lc}) +O(x-x_{\rm lc})^2, \\
c=x_{\rm lc}G'(x_{\rm lc})=1+x_{\rm lc}
\left.\left({a'\over a}-{\alpha'\over \alpha}\right)\right|_{x_{\rm lc}}
.
\end{eqnarray}
Note that $c$ is independent of $Y$, and depends only on the spacetime
curvature at the light cone. (In Fig. \ref{fig:xV} we see that all
trajectories approach the light cone with the same slope $c$.)
Integrating this we find that
\begin{equation}
\label{Qlc}
Q_\pm(x,\mp|Y|)\simeq C\left[x_0(Y),Y\right]\ |x-x_{\rm lc}|^{1\over c},
\qquad x\simeq x_{\rm lc}
\end{equation}
We see that the decay of $Q$ towards the light cone depends on $c$.
When the metric is flat we have $c=1$. This combines with a factor
$(x-x_{\rm lc})$ arising elsewhere in the integrals to give a finite
discontinuity at the light cone. However, if $c<1$ the decay is faster
and particles peeling off the light cone make no contribution at the
light cone, giving a continuous stress-energy and particle
current. Typically we find that $c\simeq 0.8$ in our examples when
gravity comes into play. Therefore gravity makes the stress-energy
tensor continuous at the light cone without the correlation between
type I and type II particles that was required for this in flat
spacetime. On the other hand, generically the stress-energy is not
analytic at the light cone because of the non-integer power of
$|x-x_{\rm lc}|$, but we could have a situation where $c^{-1}$ is an
integer and then a suitable arrangement of the particle distributions
could render an analytic stress-energy tensor. We have not been able
to construct any particular example.

\subsection{Properties of the distribution function}

Vlasov matter is considered to be a good matter model both in
Newtonian gravity and in general relativity because it does not
develop singularities in flat spacetime. For massive particles it has
been proved that singularities are also absent in the Vlasov-Poisson
system even for large initial data \cite{Pfaf92}, and for small data
in the Einstein-Vlasov system \cite{ReRe92}.

Even though it is clear that we cannot apply this second result to
massless particles, it is interesting to compare the assumptions of
that theorem with the properties of our solutions, because we can also
construct solutions which are arbitrarily close to flat spacetime. The
main assumptions of the theorem are that 1. the initial data has
compact support in both momentum space and physical space, 2. $f$ is
small, and in particular bounded, 3. $f$ is $C^1$ and the metric is
$C^2$, and 4. the spacetime has a regular center.

Using only type I particles, our metric solutions are generically only
$C^1$ at the light cone, but we believe that correlating type I with 
type II particles solutions can be made $C^2$ there.
We clearly have a regular center before the formation
of the singularity, so we only need to analyze the boundedness of $f$
and the compactness of its support at the initial time.

Our CSS solutions are infinitely extended in space, but we can match
initial data for a CSS solution for $r<r_0$ smoothly to initial data
of compact support in space. As long as $r_0>x_{\rm lc}(-t)$, the
domain of dependence of the CSS part of the data includes the
singularity. We then have a solution with compact support in space at
the initial time which is nevertheless CSS in a central region
including the singularity.

We choose a function $k(Y,Z)$ that has support only in a neighborhood of
$(Y_0,Z_0)$ not including $Y=0$ or $Z=0$. For given $t<0$ and $r$, 
$f(t,r,w,F)$ then has support in momentum space only in a neighborhood 
of the two points $(w_\pm (t,r,Y_0,Z_0),F_\pm(t,r,Y_0,Z_0))$. The two 
signs correspond to ingoing and outgoing particles, for $x<x_{\rm lc}$.
For $x>x_{\rm lc}$, only the positive sign applies. We find immediately
that
\begin{equation}
\label{FtrYZ}
F_\pm(t,r,Y_0,Z_0)=Z_0 \, t^2 Q_\pm{}^2\left(x,Y_0\right),
\end{equation}
which is finite for finite $r$. Because angular momentum is conserved
along particle trajectories, we conclude that the support of $f$ in
angular momentum is always compact.

$w_\pm(t,r,Y_0,Z_0)$ is the appropriate solution $w$ of 
\begin{equation}
\label{YtrwF}
Y=\frac{\alpha(x)}{x}\sqrt{1+\frac{(rw)^2}{F}}
+a(x)\frac{rw}{\sqrt{F}},
\end{equation}
for fixed $r$, $t$, $Y=Y_0$ and $F=F_\pm(t,r,Y_0,Z_0)$. This is also
finite. However, now $w$ is not a constant of motion and therefore
the support in $w$ does not stay constant. The typical time evolution
of the radial momentum $w$ of a type I particle is the following: an
ingoing particle starting with certain negative $w$ accelerates
towards the center, reaching a maximum modulus, and then decelerates
having a turning point and being ejected off the light cone with
positive $w$.  Eqs. (\ref{FtrYZ}) and (\ref{YtrwF}) together show us
that along lines of constant $x$, the radial momentum $w$ of the
particles that happen to be there at time $t$ is independent of
$t$. (The value of $F$ of those particles even decays as $F^2$. Note
that at each $t$ these are different particles.) Therefore the support
in radial momentum remains compact during the evolution.

We now show that in our CSS solutions $f(t,r,w,F)$ can not be finite
on the light cone $r=x_{\rm lc}(-t)$ at zero particle momentum
$w=F=0$. Consider again a function $k(Y,Z)$ that has support only in a
small neighborhood of $(Y_0,Z_0)$. Clearly if $f$ in a CSS solution is
bounded at one value of $t<0$, it is bounded for all $t<0$. Therefore
we consider a fixed value $t<0$. In order to find a limit in $(r,w,F)$
space in which
\begin{equation}
f(t,r,w,F)=\frac{1}{F}k(Y,Z)
\end{equation}
blows up, we need to find a limit in which $F\to 0$ while $Y\to Y_0$
and $Z\to Z_0$ simultaneously. Consider therefore the limit
\begin{equation}
F\to 0, \qquad 
w\to A(-t)^{-1}F^{1\over 2}, \qquad 
x \to x_{\rm lc} + BF^{c\over 2}, 
\end{equation}
where $c$ is the constant defined in (\ref{Qlc}), and $A$ and $B$ are
real constants that will be determined. From (\ref{YtrwF}) we find
that in this limit
\begin{equation}
Y\to\alpha(x_{\rm lc})\left[\left(A^2+x_{\rm lc}^{-2}\right)^{1\over
2}+A\right]\equiv Y_0(A)
\end{equation}
and therefore
\begin{equation}
Z\to C[x_0(Y_0),Y_0]^{-2}|B|^{-{2\over c}}\equiv Z_0(A,B).
\end{equation}
We can therefore always arrange the required limit for any values of
$Y_0>0$ and $Z_0$ by a suitable choice of the constants $A$ and $B$.
We have therefore shown that $f$ is infinite on the light cone at zero
momentum unless it is identically zero. From the fact that the
stress-energy tensor and the particle current are both finite if $\bar
k(y)=\int k(Y,Z) dZ$ exists, this blowup is not a physical
problem. Furthermore, for the ansatz of $k(Y,Z)$ with compact support
in $Y$ and $Z$, bounded away from $Y=0$ and $Z=0$, $f$ is finite and
can be made arbitrarily small (everywhere but at the light
cone at zero momentum).

\section{Conclusions}

Type II critical phenomena, in which the black hole mass vanishes as a
power of distance from the black hole threshold, have been found (in
some region of parameter space) for almost all Einstein-matter systems
in spherical symmetry. These include real and complex scalar fields
with arbitrary potential terms, conformal couplings, and coupling to a
Maxwell field, perfect fluids, sigma models and Yang-Mills fields. The
only exception seems to be the spherically symmetric Einstein-Vlasov
system. This raises the question what distinguishes this system from
the other ones, and the wider question if the existence of type II
critical phenomena is the rule or the exception.

Critical phenomena at the black hole threshold require the existence
of a critical solution. This is a solution that has precisely one
unstable linear perturbation mode, with the additional property that a
fully nonlinear evolution starting with a finite amplitude of this
perturbation mode results in a black hole, while a finite amplitude of
the opposite sign results in dispersion (or another outcome, such as a
star or a naked singularity).  In type II critical phenomena the
critical solution is also self-similar, either continuously (CSS) or
discretely (DSS). As a first step into understanding the absence of
type II critical phenomena we have therefore investigated the
existence and regularity of CSS solutions. 

The rest mass of the particles of the collisionless matter introduces
a scale into the coupled Vlasov and Einstein equations, which could be
incompatible with exact self-similarity. However, there are many
examples in other matter models where such a scale can be treated as a
small perturbation in a class of solutions that are asymptotically
self-similar. Type II critical phenomena are then not affected by the
presence of the scale in the field equations. We have therefore focussed
in this paper on spherically CSS solutions with a regular center of the
Einstein-Vlasov system with {\it massless} particles, hoping to
generalize these later to asymptotically CSS solutions with massive
particles. 

The main result of this paper is the explicit construction of a family
of spherically symmetric CSS solutions with massless particles that is
parameterized by an arbitrary function of two conserved quantities
$k(Y,Z)$. By function counting we have constructed the most general
spherically symmetric CSS solution with a regular center before the
formation of the singularity, but there may be particular
solutions that we have overlooked. We have also assumed the presence
of certain cutoffs for very low and very large $Y,Z$, in order to
avoid divergences.

In order to make this result more transparent, we have also rederived
the well-known general static spherically symmetric solution,
which is parameterized by an arbitrary function $h(E,F)$ of the
conserved particle energy $E$ and angular momentum $F$. In both cases,
static and CSS, we have first found the general solution for Vlasov
test particles on a fixed spacetime of that symmetry, which is
parameterized by a free function of three conserved
quantities. We have then shown that when we demand that the resulting
stress-energy tensor is compatible with the symmetry, the general
solution depends only on two conserved quantities. In the static case,
these are the obvious ones $E$ and $F$, while the choice in the CSS
case is much less obvious. 

Initial data for our solutions can be given compact support in
momentum space, and can be truncated in space without affecting a
central CSS region that includes the usual CSS singularity. However,
the Vlasov function $f(x^\mu,p^\mu)$ diverges as $p^\mu\to 0$ on the
past light cone of the singularity. This divergence is integrable, so
that both the stress-energy and the particle current are finite
everywhere in spacetime except at the CSS singularity. Furthermore,
our solutions can be constructed arbitrarily close to flat
spacetime. 

The stress-energy tensor and particle current are less differentiable
at the light cone than elsewhere. The stress-energy tensor of a
generic test particle distribution is discontinuous at the light cone,
but by imposing a relation between $k(Y,Z)$ and $k(-Y,Z)$ it can be
made $C^0$. When the Vlasov matter is coupled to gravity, we gain one
order of differentiability: the stress-energy tensor is now
generically $C^0$, and can be made $C^1$. The metric is generically
$C^1$ at the light cone, and can be made $C^2$. (Naively, one would
expect the metric to be two orders up from the stress-energy, but in
polar-radial coordinates the two metric coefficients $a$ and $\alpha$
can be determined from Einstein equations that contain only first
derivatives of $a$ and $\alpha$.)

Because our CSS solutions can be constructed with arbitrarily weak
curvature, their curvature singularity is essentially kinematic:
particles are ``aimed'' at the spacetime point where the singularity
will occur, rather than being focussed by gravity. In this context,
the relation between regularity at the past light cone and the coupling
to gravity is worth commenting on. Any spherically symmetric CSS
solution of a massless scalar test field that is regular at the center 
for $t<0$ is necessarily singular at the past light cone. Coupling the
scalar field to gravity, there exists an isolated strong field
solution which is analytic both at the center and the light cone. The
same is true for other field theories and a for a perfect fluid. By
contrast, coupling collisionless matter to gravity adds one order of
differentiability at the light cone to all matter configurations, but
no solution can be analytic at the light cone. This behavior is
mathematically more similar to the behavior of the preferred scalar
field solution at its {\it future} light cone.

We shall consider the implications of our results for type II critical
phenomena in detail elsewhere. However, we have seen that the
spacetime metric depends on the free function $k(Y,Z)$ only through
the integral $\bar k(Y)=\int k(Y,Z)dZ$. This means that there are
infinitely many matter configurations that give rise to the same
spacetime. It is clear that a similar result will hold for the
linearized perturbations of these solutions. Therefore there will be
an infinite number of linear perturbation modes with the same
eigenvalue $\lambda$. None of our solutions can therefore have a
single growing mode. This seems to be related to the fact that
Einstein-Vlasov is not a field theory. (While the phase space of the
spherically symmetric scalar field consists of pairs $\phi(r)$,
$\dot\phi(r)$, the phase space of the spherically symmetric
collisionless matter consists of functions $f(r,w,F)$: this phase
space is much bigger.)

On the other hand, due to the fact that $f$ is conserved along
particles trajectories, we cannot expect to get a very close approach
to a self-similar solution during the evolution of initial data with
bounded $f$. We have seen that in the self-similar solutions $f$ is
unbounded for low momentum near the light cone, while numerical
simulations typically work with finite $f$.

Any, or both, of those two reasons could explain why the collapse
simulations that have been carried out did not find any sign of type II
critical phenomena. We must not forget, however, that those simulations
worked with massive particles, while here we have assumed massless
particles.

\begin{acknowledgments}
We would like to thank H\aa kan Andreasson and Alan Rendall for
interesting conversations and comments. This research was funded in
part by EPSRC grant GR/N10172.
\end{acknowledgments}

\bibliography{critvlasov4}

\end{document}